\documentclass[preprint,showpacs,amsmath,amssymb,aps,prd,nofootinbib]{revtex4}
\usepackage{epsfig,graphicx,color,appendix,fge}
\begin{document}
\begin{flushright}
{ CAU-THEP-20-10}
\end{flushright}
\def\CP{{\it CP}~}
\def\cp{{\it CP}}
\title{\mbox{}\\[10pt]
Challenge to Anomalous Phenomena\\
in Solar Neutrino}

\author{Y. H. Ahn}
\affiliation{Department of Physics, Chung-Ang University, Seoul 06974, Korea.}
\email{axionahn@naver.com}




\begin{abstract}
 We suggest a would-be solution to the solar neutrino tension why solar neutrinos appear to mix differently from reactor antineutrinos, in theoretical respect. To do that, based on an extended theory with light sterile neutrinos added we derive a general transition probability of neutrinos born with one flavor tuning into a different flavor. 
 Three new mass-squared differences are augmented in the extended theory: two $\Delta m^2_{\rm ABL}\lesssim{\cal O}(10^{-11})\,{\rm eV}^2$ optimized at astronomical-scale baseline (ABL) oscillation experiments and one $\Delta m^2_{\rm SBL}\sim{\cal O}(1)\,{\rm eV}^2$ optimized at reactor short-baseline (SBL) oscillation experiments. With a so-called composite matter effect that causes a neutrino flavor change via the effects of sinusoidal oscillation including the Mikheyev-Smirnov-Wolfenstein matter effect, we find that the value of $\Delta m^2$ measured from reactor antineutrino experiments can be fitted with that from the $^8$B solar neutrino experiments for roughly $\Delta m^2_1\lesssim10^{-13}\,{\rm eV}^2$ and $\Delta m^2_2\simeq{\cal O}(10^{-11})\,{\rm eV}^2$. Nonetheless, we find that the current data (solar neutrino alone) is not precise enough to test the proposed scenario.
Future precise measurements of $^8$B and $pep$ solar neutrinos may confirm and/or improve the value of $\Delta{m}^2_2$. 
\end{abstract}

\maketitle %
\section{Introduction}
 One of the great discoveries in particle physics is the experimental evidence of neutrino oscillations, implying that neutrinos are massive particles and that
the three flavor neutrinos $\nu_e, \nu_\mu, \nu_\tau$ 
are mixtures of neutrinos with definite masses $\nu_i$ (with $i=1,2,...$) which are identified based on the charged particles (electron, muon, tau) produced via weak interaction\,\cite{PDG}.
Despite the successful description of the properties of three known neutrino species through decades of experimentation, we are faced with a growing list of anomalous phenomena -- experimentally unexpected results that conflict with the three active neutrino oscillation standard framework ($3\nu$SF\,\cite{PDG}): (i) the so-called ``short-baseline (SBL) anomalies"\,\cite{Mention:2011rk, gallium1, gallium2, Athanassopoulos:1995iw} (including MiniBooNE data\,\cite{Aguilar-Arevalo:2018gpe}), anomalous results measured in several experiments at distance less than $1$ km, 
and (ii) the  so-called ``solar neutrino tension"\,\cite{Abe:2016nxk, Capozzi:2018ubv, Esteban:2018azc} (see also Refs.\,\cite{Collaboration:2011nga, Agostini:2017ixy}), a discrepancy between the oscillation parameter determined in solar neutrino experiments and the one measured in reactor neutrino oscillation experiments. 
Although the solar neutrino tension (below $2\sigma$ uncertainties) seems small, the discrepancy has been the long-standing tension in the neutrino physics of $3\nu$SF\,\cite{Esteban:2018azc, Abe:2016nxk, Capozzi:2018ubv}. In fact, it is well known that 
none of the $^8$B measurements performed by Sudbury Neutrino Observatory (SNO)\,\cite{Collaboration:2011nga}, Super-Kamiokande (SK)\,\cite{Abe:2016nxk}, and Borexino\,\cite{Agostini:2017ixy} has shown any evidence of the low energy spectrum turn-up expected in the standard Mikheyev-Smirnov-Wolfenstein (MSW)-the large mixing angle, LMA, solution for the value of mass-squared difference $\Delta m^2$ favored by KamLAND\,\cite{Eguchi:2002dm, Araki:2004mb, Gando:2013nba}. Moreover, the recent observation of low-energy $^8$B solar neutrino flux (as low as $\sim3.5$ MeV) by SK\,\cite{Abe:2016nxk} has marked the raise of the solar neutrino tension. The anomalous phenomena (the SBL anomalies plus solar neutrino tension) could be due to some unknown physical phenomena that are sensitive to SBL oscillations and solar neutrino oscillations, or else unlikely statistical fluctuations in the current data, or experimental errors.
If such anomalous phenomena point to potential problems with the $3\nu$SF predictions, they may provide a new level of importance as possible routes to ``new neutrino physics", an extended theory for phenomena unexplained by the $3\nu$SF.

In theoretical respects, the most straightforward interpretation of those anomalous phenomena could be neutrino oscillations with new parameters.
Problem is, in the $3\nu$SF\,\cite{PDG}, there are only six oscillation parameters: two mass-squared differences $\Delta m^2_{\rm Sol}$ and $\Delta m^2_{\rm Atm}$, three mixing angles $\theta_{23}, \theta_{13}, \theta_{12}$, and one Dirac CP phase $\delta_{CP}$. These parameters are not sufficient to fit the anomalous phenomena, indicating that the current theoretical understanding of neutrino oscillation may not be complete. 
If the anomalous phenomena are interpreted in terms of neutrino oscillations, it may suggest the presence of other types of light neutrinos called sterile neutrinos that do not have weak interactions\,\footnote{The authors in Ref.\,\cite{Mohapatra:2005wk} have extended the Pontecorvo-Maki-Nakagawa-Sakata (PMNS) mixing matrix to interpret the SBL anomalies as neutrino oscillations. While, instead of the way of extending the PMNS matrix, in order to introduce new oscillation parameters the authors in Ref\,.\cite{Ahn:2019jbm} have parameterized with unitary condition in a way that a diagonal form of $2\times2$ partitioned matrix holding the $3\times3$ PMNS mixing matrix is linearly multiplied by a $6\times6$ mixing matrix of active to sterile neutrinos.}. 
An attempt to explain the SBL anomalies has been proposed in a framework of neutrino oscillation\,\cite{Ahn:2019jbm}. Moreover, in the same framework for neutrino oscillation the experimental results on high energy neutrinos released from IceCube\,\cite{icecube} could be interpreted\,\footnote{The track-to-shower ratio of a cosmic neutrino\,\cite{Palladino:2015zua} can give a new oscillation curve as a signal dependent on neutrino flight length if the neutrino mixing parameters of 3$\nu$SF, initial flavor composition, and tiny
mass splittings are given as inputs.} as new oscillation effects\,\cite{Ahn:2019jbm, Ahn:2016hhq}.

The goal of this work is to study for a theoretical understanding of the solar neutrino tension why solar neutrinos at SNO, SK, and Borexino experiments appear to mix differently from reactor antineutrinos at KamLAND, despite that both neutrinos are sensitive to the same oscillation parameters in the $3\nu$SF.
To do that,  first, based on an extended theory with light sterile neutrinos added\,\cite{Ahn:2019jbm} we derive a general transition probability between the massive neutrinos that a flavor eigenstate $\nu_\alpha$ becomes flavor eigenstate $\nu_\beta$ with $\alpha,\beta=e,\mu,\tau$, that can also have a potential for explaining both the anomalous phenomena and ultra-high energy neutrino events at IceCube, simultaneously\,\footnote{However, in this paper, we will not study phenomenological interpretations of the SBL anomalies and astronomical neutrino data at IceCube, see Refs.\,\cite{Ahn:2019jbm, Ahn:2016hhq}.}. Second, we re-examine the MSW matter effects in our theoretical framework and suggest a solution to the solar neutrino tension with a so-called composite matter effect that causes a neutrino flavor change with new oscillatory terms containing $\Delta m^2_{\rm ABL}\lesssim{\cal O}(10^{-11})\,{\rm eV}^2$ optimized at astronomical-scale baseline (ABL) ($\gtrsim L_{es}=149.6\times10^6$ km, earth-sun distance) oscillation experiments together with the MSW matter effect. An important point is that, contrary to the MSW effect\, \cite{Wolfenstein:1977ue, Mikheev:1986gs} that causes a change in the flavor content of a neutrino but without sinusoidal oscillation, the so-called composite matter effect causes a neutrino flavor change via the effects of sinusoidal oscillation, as well as the MSW matter effect.

This work is organized as follows. In section II we provide an introduction to the model setup, masses and mixings. In section III we compute a general transition probability for three flavor neutrinos with their three light sterile neutrino pairs, 
subsequently, in section III-A we investigate possible mass orderings and show how additional oscillation parameters could be constrained by cosmological data (the sum of active neutrino masses) and the effective neutrino mass in both $\beta$-decay and neutrinoless-double-beta ($0\nu\beta\beta$)-decay experiments, and then we interpret reactor anti-neutrinos at KamLAND in the new oscillation framework in section III-B. In section IV-A, we re-examine the MSW matter effects in the extended $3\nu$SF, and we analytically study why the oscillation parameters determined in solar neutrino experiments are not in complete agreement with the measurements collected in other types of experiments in section IV-B. Conclusions are drawn in section V.

\section{Masses and Mixings}
In the basis of interaction eigenstates, $\psi_L\equiv(\nu_L~S^c_R)^T$, where active neutrinos are in the up-stairs and sterile neutrinos are in the down-stairs, the most general renormalizable Lagrangian for neutrinos reads in the charged lepton basis at low energies\,\cite{Ahn:2019jbm}
 \begin{eqnarray}
-{\cal L}_{\nu} &=& 
 \frac{1}{2} \begin{pmatrix} \overline{\nu^c_L} & \overline{S_R} \end{pmatrix} {\cal M}_{\nu} \begin{pmatrix} \nu_L \\ S^c_R  \end{pmatrix} +\frac{g}{\sqrt{2}}W^-_\mu\overline{\ell_L}\gamma^\mu\,\nu_L+\text{h.c.}+\frac{g}{2\cos\theta_W}Z_\mu\bar{\nu}_L\gamma^\mu\nu_L\,,
  \label{lag}
 \end{eqnarray}
 where $g$ is the $SU(2)$ coupling constant, $\theta_W$ is the Weinberg angle, $\ell=(e, \mu, \tau)$, $\nu_L=(\nu_e, \nu_\mu, \nu_\tau)$, and $S_R=(S_1, S_2,...S_n)$. The light neutral fermions $S_\alpha$ do not take part in the standard weak interaction and thus are not excluded by LEP results, while the number of active neutrinos that are coupled with the $W^{\pm}$ and $Z$ bosons is $N_{\nu}=2.984\pm0.008$\,\cite{Beringer:1900zz}. After electroweak symmetry breaking, Eq.\,(\ref{lag}) describes $3\times n$ Majorana neutrinos. In the case of $n=3$ sterile neutrinos, the $6\times6$ Majorana neutrino mass matrix is\,\footnote{Consider a Lagrangian extended by another type of Majorana neutrinos $N_R$ whose masses are much larger than the scale of electroweak symmetry breaking scale. After integrating out such heavy degrees of freedom, small left-handed Majorana neutrino masses, $M_L\simeq-m^T_DM^{-1}_Rm_D$, can be naturally generated via seesaw mechanism\,\cite{Minkowski:1977sc} where $m_D$ and $M_R$ are Dirac and Majorana neutrino masses, see {\it e.g.} Ref.\,\cite{Ahn:2016hhq, Ahn:2018nfb}.}
 \begin{eqnarray}
{\cal M}_{\nu} = \begin{pmatrix} M_L & M^T_{D}  \\ M_{D} & M_{S}  \end{pmatrix} \,,
  \label{nu_matr}
 \end{eqnarray}
which is complex and symmetric, where $M_{D}$, $M_L$, and $M_S$ are $3\times3$ mass matrices for Dirac masses, left- and right-handed Majorana masses, respectively.
Thus it can be diagonalized by a $6\times6$ matrix $W_\nu$ through basis rotations from interaction eigenstates $\psi_L$ to mass eigenstates $n_L$, that is,
 \begin{eqnarray}
 \psi_L=\begin{pmatrix} \nu_L \\ S^c_R  \end{pmatrix} \rightarrow W^\dag_\nu\begin{pmatrix} \nu_L \\ S^c_R  \end{pmatrix}\equiv n_L\,.
  \label{Wnu0}
 \end{eqnarray}
The phenomenology of Eq.\,(\ref{nu_matr}) depends on the values of the matrices $M_L$, $M_D$, and $M_S$, that is, the mass eigenvalues and mixings of $M_D$, $M_S$ and $M_L$.
We impose unitary condition to $W_\nu$ (such that satisfys the unitary condition $W_{\nu}W^\dag_\nu=W^\dag_{\nu}W_\nu=I_{6\times6}$), which preserves norm and thus probability amplitude, and choose the $6\times6$ unitary neutrino transformation matrix as\,\cite{Ahn:2016hhq, Ahn:2019jbm}
 \begin{eqnarray}
 W_\nu={\left(\begin{array}{cc}
 U_LV_1 & iU_LV_1  \\
 U_RV_2 &  -iU_RV_2 
 \end{array}\right)}V_\nu
  \label{Wnu}
 \end{eqnarray}
 where $U_L$ corresponds to the PMNS mixing matrix $U_{\rm PMNS}$, $U_R$ is an unknown unitary $3\times3$ matrix, $V_1={\rm diag}(1,1,1)/\sqrt{2}$, and $V_2={\rm diag}(e^{i\chi_1}, e^{i\chi_2}, e^{i\chi_3})/\sqrt{2}$ with $\chi_i$ being arbitrary phases.
  The $6\times6$ unitary mixing matrix $V_\nu$ forms a bridge between active and sterile neutrinos:
  \begin{eqnarray}
 V_\nu= {\left(\begin{array}{cccccc}
 e^{i\phi_1}\,\cos\theta_1 & 0 & 0 &  -e^{i\phi_1}\,\sin\theta_1 & 0 & 0 \\
 0 & e^{i\phi_2}\,\cos\theta_2 & 0 & 0 & -e^{i\phi_2}\,\sin\theta_2 & 0 \\
  0 & 0 & e^{i\phi_3}\,\cos\theta_3 & 0 & 0 & -e^{i\phi_3}\,\sin\theta_3 \\
 e^{-i\phi_1}\,\sin\theta_1 & 0 & 0 &  e^{-i\phi_1}\,\cos\theta_1 & 0 & 0 \\
 0 & e^{-i\phi_2}\,\sin\theta_2 & 0 & 0 & e^{-i\phi_2}\,\cos\theta_2 & 0 \\
  0 & 0 & e^{-i\phi_3}\,\sin\theta_3 & 0 & 0 & e^{-i\phi_3}\,\cos\theta_3 
 \end{array}\right)}\,.
 \label{nu_mix}
 \end{eqnarray}
In the mass eigenstates $\nu_1, \nu_2, \nu_3$, $S^c_1,S^c_2,S^c_3$  basis the Hermitian matrix ${\cal M}_\nu{\cal M}^\dag_\nu$ can be diagonalized as a real and positive $6\times6$ mass-squared matrix by the unitary transformation $W_\nu$ of Eq.\,(\ref{Wnu})
 \begin{eqnarray}
   &W^T_\nu\,{\cal M}_\nu{\cal M}^\dag_\nu\,W^\ast_\nu\equiv{\rm diag}(m^2_{\nu_1}, m^2_{\nu_2}, m^2_{\nu_3}, m^2_{s_1},m^2_{s_2}, m^2_{s_3})\nonumber\\
   &= V^T_\nu{\left(\begin{array}{cc}
 |\hat{M}|^2+|\hat{M}||\delta|+\frac{1}{2}(|\hat{M}_L|^2+|\hat{M}_S|^2) &  \frac{i}{2}(|\hat{M}_S|^2-|\hat{M}_L|^2)  \\
 -\frac{i}{2}(|\hat{M}_S|^2-|\hat{M}_L|^2) &  |\hat{M}|^2-|\hat{M}||\delta|+\frac{1}{2}(|\hat{M}_L|^2+|\hat{M}_S|^2) 
 \end{array}\right)}V^\ast_\nu\,,
  \label{eff_nu_mass}
 \end{eqnarray}
 where a real positive diagonal matrix $\hat{M}=U^T_R\,M_{D}\,U_L$$={\rm diag}(m_1, m_2, m_3)$ and two complex diagonal matrices $\hat{M}_{L}=U^T_LM_{L}U_L$ and $\hat{M}_{S}\equiv U^T_RM_{S}U_R$ are used. 
In Eq.\,(\ref{eff_nu_mass}), the parameter $\delta$ is defined\,\footnote{The real positive condition of $\hat{M}=U^T_R\,M_{D}\,U_L$ derives the phase of $V_2$ to be equivalent to that of $\delta$.} as
\begin{eqnarray}
 \delta_k\equiv (\hat{M}_{L})_k+(\hat{M}^\dag_S)_k\,\quad\text{with}~\arg(\delta_k)=\chi_k\,(k=1,2,3),
  \label{}
\end{eqnarray}
where the mass-squared eigenvalues (real and positive) are given by
\begin{eqnarray}
 m^2_{\nu_k}&=& m_k^2+\frac{1}{2}(|(\hat{M}_L)_k|^2+|(\hat{M}_S)_k|^2)+\frac{m_k|\delta_k|}{\cos2\theta_k}\,,\nonumber\\
 m^2_{s_k}&=& m_k^2+\frac{1}{2}(|(\hat{M}_L)_k|^2+|(\hat{M}_S)_k|^2)-\frac{m_k|\delta_k|}{\cos2\theta_k}\,,
  \label{nu_mass}
\end{eqnarray}
and the mixing angles $\theta_k$ between active and sterile neutrino is given by 
 \begin{eqnarray}
  \tan2\theta_k=\frac{|(\hat{M}_L)_k|^2-|(\hat{M}_{S})_k|^2}{2m_k|\delta_k|}\sin2\phi_k\,.
 \end{eqnarray}
From Eq.\,(\ref{nu_mass}) we define the mass splitting for $k$-th generation  as  
  \begin{eqnarray}
  \Delta m^2_k\equiv m^2_{\nu_k}-m^2_{s_k}=2\frac{|\delta_k|\,m_{k}}{\cos2\theta_k}\,.
  \label{msplit}
 \end{eqnarray}
Here, for simplicity, we take $\phi_{k}=-\pi/4$ and assume $|(\hat{M}_S)_k|, m_k\gg|(\hat{M}_L)_k|$, or equivalently, ${\cal M}_{\nu} \simeq{\tiny\begin{pmatrix} 0 & M^T_{D}  \\ M_{D} & M_{S}  \end{pmatrix}}$ in Eq.\,(\ref{nu_matr}).
 Then the mass-squared eigenvalues of $k$-th generation of Eq.\,(\ref{nu_mass}) can be expressed in terms of $\theta_k$ and $\Delta m^2_k$ as
\begin{eqnarray}
 m^2_{\nu_k}= \frac{(1+\sin2\theta_k)^2}{4\sin2\theta_k}\Delta m^2_k\,,\qquad
 m^2_{s_k}= \frac{(1-\sin2\theta_k)^2}{4\sin2\theta_k}\Delta m^2_k\,,
  \label{nu_mass1}
\end{eqnarray}
which in turn, together with Eq.\,(\ref{msplit}), leads to
\begin{eqnarray}
 \Delta m^2_k\left\{
       \begin{array}{ll}
         <0\, \Rightarrow& \pi/2<\theta_k<3\pi/4 \\
         >0\, \Rightarrow& 0<\theta_k<\pi/4
       \end{array}
     \right..
  \label{para0}
\end{eqnarray}

The three active neutrino states emitted by weak interactions are described in terms of the mass eigenstates $\nu_k$, $S_k$ ($k=1,2,3$) and the $3\times3$ PMNS mixing matrix $U\equiv U_{\rm PMNS}$ as
 \begin{eqnarray}
   \nu_\alpha=\sum_{k=1}^3U_{\alpha k}\,n_k
  \label{nu_mass_eigen}
 \end{eqnarray}
 with the massive states 
 \begin{eqnarray}
 n_{k}=\frac{1}{\sqrt{2}}\left(\begin{array}{cc}
 \cos\theta_k-\sin\theta_k & \cos\theta_k+\sin\theta_k  \end{array}\right)\left(\begin{array}{c}
 \nu_{k} \\
 S_k \end{array}\right)\,,
  \label{nu_mass_eigen}
 \end{eqnarray}
in which the field redefinitions $\nu_k\rightarrow e^{-i\pi/4}\nu_k$ and $S_k\rightarrow e^{i3\pi/4}S_k$ have been used. 
Since the active neutrinos are massive and mixed, the weak eigenstates $\nu_{\alpha}$ (with flavor $\alpha=e,\mu,\tau$) produced in a weak gauge interaction are linear combinations of the mass eigenstates with definite masses. The charged gauge interaction in Eq.\,(\ref{lag}) for the neutrino flavor production and detection is written in the charged lepton basis as
\begin{align}
-{\cal L}_{\rm c.c.} =\frac{g}{\sqrt{2}}W^{-}_{\mu}\overline{\ell_{\alpha}}\,\frac{1+\gamma_5}{2}\gamma^{\mu}\,U_{\alpha k}\,n_{k}+ {\rm h.c.}\,.
\label{WK1}
\end{align}
 Thus
in the mass eigenstate basis the PMNS leptonic mixing matrix\,\cite{PDG} at low energies is visualized in the charged weak interaction, which is expressed in terms of three mixing angles, $\theta_{12}, \theta_{13}, \theta_{23}$, and three \cp-odd phases (one $\delta_{CP}$ for the Dirac neutrino and two $\tilde{\varphi}_{1,2}$ for the Majorana neutrino) as
 \begin{eqnarray}
  U_{\rm PMNS}=
  {\left(\begin{array}{ccc}
   c_{13}c_{12} & c_{13}s_{12} & s_{13}e^{-i\delta_{CP}} \\
   -c_{23}s_{12}-s_{23}c_{12}s_{13}e^{i\delta_{CP}} & c_{23}c_{12}-s_{23}s_{12}s_{13}e^{i\delta_{CP}} & s_{23}c_{13}  \\
   s_{23}s_{12}-c_{23}c_{12}s_{13}e^{i\delta_{CP}} & -s_{23}c_{12}-c_{23}s_{12}s_{13}e^{i\delta_{CP}} & c_{23}c_{13}
   \end{array}\right)}P_{\nu}~,
 \label{PMNS}
 \end{eqnarray}
where $s_{ij}\equiv \sin\theta_{ij}$, $c_{ij}\equiv \cos\theta_{ij}$ and $P_{\nu}$ is a diagonal phase matrix what is that particles are Majorana ones. Three-flavor oscillation parameters from global fit results at the best-fit values and $(1\sigma)\,3\sigma$ confidence intervals in Ref.\,\cite{Esteban:2018azc} for normal mass ordering, NO, [inverted one, IO] in Table-\ref{exp_nu}.
\begin{table}[h]
\caption{\label{exp_nu} The global fit of three-flavor oscillation parameters at the best-fit and $(1\sigma)3\sigma$ level with Super-Kamiokande atmospheric data\,\cite{Esteban:2018azc}. NO = normal neutrino mass ordering; IO = inverted mass ordering with $\Delta m^{2}_{\rm Sol}\equiv m^{2}_{\nu_2}-m^{2}_{\nu_1}$, $\Delta m^{2}_{\rm Atm}\equiv m^{2}_{\nu_3}-m^{2}_{\nu_1}$ for NO, and  $\Delta m^{2}_{\rm Atm}\equiv m^{2}_{\nu_2}-m^{2}_{\nu_3}$ for IO.}
\begin{ruledtabular}
\begin{tabular}{cccccccccccc} &$\theta_{13}[^{\circ}]$&$\delta_{CP}[^{\circ}]$&$\theta_{12}[^{\circ}]$&$\theta_{23}[^{\circ}]$&$\Delta m^{2}_{\rm Sol}[10^{-5}{\rm eV}^{2}]$&$\Delta m^{2}_{\rm Atm}[10^{-3}{\rm eV}^{2}]$\\
\hline
$\begin{array}{ll}
\hbox{NO}\\
\hbox{IO}
\end{array}$&$\begin{array}{ll}
8.61^{+(0.12)0.37}_{-(0.13)0.39} \\
8.65^{+(0.12)0.38}_{-(0.13)0.38}
\end{array}$&$\begin{array}{ll}
217^{+(40)149}_{-(28)82} \\
280^{+(25)71}_{-(28)84}
\end{array}$&$33.82^{+(0.78)2.45}_{-(0.76)2.21}$&$\begin{array}{ll}
49.7^{+(0.9)2.5}_{-(1.1)8.8} \\
49.7^{+(0.78)2.4}_{-(0.75)8.5}
\end{array}$&$7.39^{+(0.21)0.62}_{-(0.20)0.69}$
 &$\begin{array}{ll}
2.525^{+(0.033)0.097}_{-(0.031)0.094} \\
2.512^{+(0.033)0.094}_{-(0.031)0.099}
\end{array}$ \\
\end{tabular}
\end{ruledtabular}
\end{table}

On the other hand, as shown in Ref.\,\cite{Abe:2016nxk}, in the $3\nu$SF there is a clear discrepancy for mass-squared difference at $\lesssim2\sigma$ C.L.: the value $\Delta m^2_{\rm KL}$ preferred by KamLAND\,\cite{Gando:2013nba} is somewhat higher than the one $\Delta m^2_{\odot}$ from solar experiments\,\cite{Cleveland:1998nv, Abe:2016nxk, Aharmim:2011vm, Agostini:2017ixy, Fukuda:2001nj}, while their mixing angles are overlapped, which is as follows at $2\sigma$ C.L. for $\theta_{13}[^\circ]=8.51^{+0.27}_{-0.28}$,
 \begin{eqnarray}
  &\Delta m^2_{\odot}=4.85^{+1.33}_{-0.59}\times10^{-5}\,{\rm eV}^2\,,\qquad\qquad\theta_{\odot}[^\circ]=33.71^{+0.86}_{-0.87}\,,\nonumber\\
  &\Delta m^2_{\rm KL}=7.49^{+0.19}_{-0.18}\times10^{-5}\,{\rm eV}^2\,,\qquad\qquad\theta_{\rm KL}[^\circ]=33.65^{+0.80}_{-0.75}\,.
 \label{sol_kam}
 \end{eqnarray}
 The oscillation parameter $\Delta{m}^2$ determined in solar neutrino experiments is not in complete agreement with the measurements collected in other types of reactor and accelerator experiments: the so-called ``solar neutrino tension".
 
\section{Mass-Induced neutrino Oscillations in vacuum}
The parameterization of Eq.\,(\ref{Wnu}) as a mixing matrix of active to sterile neutrinos, leading to interferences between active and sterile neutrinos, has additional oscillation parameters $(\Delta m^2, \theta)$ that trigger new oscillation effects in charge of both explanations of SBL anomalies and solar neutrino tension, as well as ultra-high energy neutrino events at IceCube. 
Such additional oscillation parameters $\Delta m^2$ and $\theta$ appear in the expression of active neutrino masses, modifying the standard form of transition probability of the $3\nu$SF, that are up to the eyes in the cosmological data (the sum of active neutrino masses), three active neutrino oscillation data, and effective neutrino masses of both $\beta$-decay and $0\nu\beta\beta$-decay experiments. 

Let us first bring out a transition probability of new oscillations with the help of the neutrino mixing matrix Eq.\,(\ref{Wnu}). The transition probability between the massive neutrinos that a neutrino eigenstate $\nu_a$ becomes an eigenstate $\nu_b$ follows from the time evolution of mass eigenstates as
\begin{eqnarray}
P_{\nu_a\rightarrow\nu_b}(W_\nu, L, E)&=&\Big|(W_\nu\,e^{-i\frac{\hat{\cal M}^2_\nu}{2E}L}W^\dag_\nu)_{ab}\Big|^2\,,
\label{proba}
\end{eqnarray}
where $a,b=e,\mu,\tau,s_1,s_2,s_3$, $L$ is the distance between the neutrino detector and the neutrino
source, $E$ is the neutrino energy, and $\hat{\cal M}_\nu\equiv W^T_\nu{\cal M}_\nu W_\nu$.
We are interested in the flavor transition between the active neutrinos $\nu_e,\,\nu_\mu,\,\nu_\tau$. From Eq.\,(\ref{proba}) the flavor transition probability between the active neutrinos $\nu_{e,\mu,\tau}$ can be generically expressed in terms of the oscillation parameters $\theta$, $\Delta m^2$, $L$, $E$, and the mixing components of the $3\times3$ PMNS matrix $U_{\alpha i}$ as
\begin{eqnarray}
P_{\nu_\alpha\rightarrow\nu_\beta}&=&\delta_{\alpha\beta}-\sum^3_{k=1}|U_{\alpha k}|^2|U_{\beta k}|^2\sin^2\Big(\frac{\Delta m^2_k}{4E}L\Big)\cos^22\theta_k\nonumber\\
&-&\sum_{k>j}{\rm Re}\big[U^\ast_{\beta k}U_{\beta j}U^\ast_{\alpha j}U_{\alpha k}\big]\sin^2\tilde{\Delta}_{kj}
+\frac{1}{2}\sum_{k>j}{\rm Im}\big[U^\ast_{\beta k}U_{\beta j}U^\ast_{\alpha j}U_{\alpha k}\big]\sin\tilde{\Delta}_{kj}\,,
\label{osc01}
\end{eqnarray}
where 
\begin{eqnarray}
\sin^2\tilde{\Delta}_{kj}&=&(1-\sin2\theta_k)\Big\{\sin^2\Big(\frac{\Delta m^2_{kj}}{4E}L\Big)(1-\sin2\theta_j)
+\sin^2\Big(\frac{\Delta Q^2_{kj}}{4E}L\Big)(1+\sin2\theta_j)\Big\}\nonumber\\
&+&(1+\sin2\theta_k)\Big\{\sin^2\Big(\frac{\Delta Q^2_{jk}}{4E}L\Big)(1-\sin2\theta_j)
+\sin^2\Big(\frac{\Delta S^2_{kj}}{4E}L\Big)(1+\sin2\theta_j)\Big\},\label{osc02}
\end{eqnarray}
\begin{eqnarray}
\sin\tilde{\Delta}_{kj}&=&(1-\sin2\theta_k)\Big\{\sin\Big(\frac{\Delta m^2_{kj}}{2E}L\Big)(1-\sin2\theta_j)
+\sin\Big(\frac{\Delta Q^2_{kj}}{2E}L\Big)(1+\sin2\theta_j)\Big\}\nonumber\\
&-&(1+\sin2\theta_k)\Big\{\sin\Big(\frac{\Delta Q^2_{jk}}{2E}L\Big)(1-\sin2\theta_j)
-\sin\Big(\frac{\Delta S^2_{kj}}{2E}L\Big)(1+\sin2\theta_j)\Big\},\label{osc03}
\end{eqnarray}
with $\Delta m^2_{kj}\equiv m^2_{\nu_k}-m^2_{\nu_j}$, $\Delta S^2_{kj}\equiv m^2_{s_k}-m^2_{s_j}$, and $\Delta Q^2_{kj}\equiv m^2_{\nu_k}-m^2_{s_j}$. 
In the model the mixing parameters $\theta$ and $\Delta m^2$ are determined by nature, so experiments should choose $L$ and $E$ to be sensitive to oscillations through a given $\Delta m^2$.
As expected, in the limit of $m_{s_i}\rightarrow m_{\nu_i}$ and $\theta_i\rightarrow0$ ($i=1,2,3$), the new transition probability of Eq.\,(\ref{osc01}) is recovered to the standard form of that for three neutrinos in vacuum\,\cite{PDG}.

\subsection{Possible mass orderings}
In the new oscillation framework of Eq.\,(\ref{osc01}) there appear three additional oscillation parameter sets ($\Delta m^2_i, \theta_i$) in addition to the six standard oscillation parameters.
To resolve a tension between the mass-squared differences of solar and reactor neutrinos in Eq.\,(\ref{sol_kam}), as well as to accommodate an eV sterile neutrino for a possible solution to the SBL anomalies\,\cite{Ahn:2019jbm} and ultra-high energy neutrino events at the IceCube detector\,\cite{Ahn:2019jbm, Ahn:2016hhq, ice_ref}, we assume 
\begin{eqnarray}
|\Delta m^2_1|\sim \frac{4\pi E}{L_{\rm ABL1}}\ll|\Delta m^2_2|\sim \frac{4\pi E}{L_{\rm ABL2}}\ll\Delta m^2_{\rm Sol}\ll\Delta m^2_{\rm Atm}\ll |\Delta m^2_3|\sim\frac{4\pi E}{L_{\rm SBL}}
  \label{assu}
\end{eqnarray}
together with $|\cos2\theta_{1(2)}|\approx1$ and a sizable $\theta_3$ from Eq.\,(\ref{msplit}), where $L_{\rm ABL}$ and $L_{\rm SBL}$ stand for astronomical-scale baselines ($\gtrsim L_{es}$) and short-baselines ($\lesssim1$km), respectively. Hence Eq.\,(\ref{assu}) means
 the effects of the pseudo-Dirac neutrinos for the first and second generation characterized by $\Delta m^2_{1(2)}$ can be detected through ABL oscillation experiments without damaging the three-active neutrino oscillation experimental data, while that for the third generation characterized by $\Delta m^2_{3}$ can be measured through SBL oscillation experiments\,\cite{Ahn:2019jbm}. Moreover, Eq.\,(\ref{assu}) indicates that the mass splittings $\Delta m^2_{1(2)}$ should be constrained by results in reactor and accelerator based neutrino experiments, such as the results of reactor experiments (optimized at oscillation lengths $L\sim$ km with $\Delta m^2_{31}\sim2.5\times10^{-3}\,{\rm eV}^2$ and $E_{\bar{\nu}_e}\sim$ MeV) and solar neutrino oscillation experiments (optimized at $L\sim{\cal O}(10\sim100)$ km with $\Delta m^2_{21}\sim7.5\times10^{-5}\,{\rm eV}^2$ and $E_{\bar{\nu}_e}\sim$ MeV), since they can modify the reactor angle $\theta_{13}$, atmospheric mixing angle $\theta_{23}$, and the LMA solution $\theta_{12}$. Whereas\,\footnote{Recent cosmological data have a tendency toward disfavoring an excess of radiation beyond the three neutrinos and photons: considering the Big Bang Nucleosynthesis (BBN) and cosmic microwave background (CMB) limits to $\Delta N^{\rm eff}_\nu<0.2$ at $95\%$ CL\,\cite{Cyburt:2015mya}. The SBL anomalies including MiniBooNE data may indicate the existence of eV-mass sterile neutrino, while present cosmological data coming from CMB+LSS, and BBN do not prefer extra fully thermalized sterile neutrino in the eV-mass range since they violate the hot dark matter limit on the neutrino mass. It can be realized by requiring sterile neutrinos do not or partially equilibrium at the BBN epoch when the initial lepton asymmetry is large\,\cite{BBN, Abazajian:2004aj, Hannestad:2012ky, Chu:2006ua, Barbieri:1990vx}. Especially in Ref.\,\cite{Hannestad:2012ky} showed quantitatively the amount of thermalization as a function of neutrino parameters (mass splitting, mixing, and large lepton asymmetry) but with a loophole for eV sterile neutrinos to be compatible with the extra energy density preferred by CMB + LSS.}  the mass splitting $\Delta m^2_{3}\sim{\cal O}(1)\,{\rm eV}^2$ can be constrained by SBL ($\lesssim1$km) oscillation experiments (and possibly long baseline oscillation experiments) as shown in Ref.\,\cite{Ahn:2019jbm}.

Considering the hierarchy of mass splittings Eq.\,(\ref{assu}), there are two possible neutrino mass spectra\,\footnote{Here the possibilities of mass ordering via the sign of $\Delta m^2_{1(2)}$ in Eq.\,(\ref{para0}) are not considered.} by taking an order of eV mass splitting $\Delta m^2_3=m^2_{\nu_3}-m^2_{s_3}<0$ into account as in Ref.\,\cite{Ahn:2019jbm}: (i) the normal mass ordering   $m^2_{\nu_1}\approx m^2_{s_1}< m^2_{s_2}\sim m^2_{\nu_2}<m^2_{\nu_3}\ll m^2_{s_3}$ (NO) and (ii) the inverted mass ordering $m^2_{\nu_3}<m^2_{\nu_1}\approx m^2_{s_1}<m^2_{s_2}\sim m^2_{\nu_2}\ll m^2_{s_3}$ (IO), because of the observed hierarchy $\Delta m^{2}_{\rm Atm}\gg\Delta m^{2}_{21}$ and the requirement of a MSW resonance for solar neutrinos $\Delta m^2_{21}>0$, see Eq.\,(\ref{mat1_12}). 
Since oscillation experiments are insensitive to the absolute scale of neutrino masses, we study how the new mixing parameters ($\theta_k$, $\Delta m^2_{k}$) can be constrained through the sum of three active neutrinos $\sum m_\nu$.
 Cosmology is mostly sensitive to the total energy density in neutrinos, directly proportional to the sum of the neutrino masses which can be expressed in terms of $\theta_k$ and $\Delta m^2_k$, see Eq.\,(\ref{nu_mass1}). 
Together with the known mass-squared differences ($|\Delta m^2_{31}|\sim|\Delta m^2_{32}|$, $\Delta m^2_{21}$) in Table-\ref{exp_nu} of the standard oscillation form, the sum of the neutrino masses can be parameterized in terms of new parameters ($\Delta m^2_{k}$, $\theta_k$):
\begin{eqnarray}
 \sum_{k=1}^3 m_{\nu_k}(\theta_1,\Delta m^2_1)&=&m_{\nu_1}+\sqrt{m^2_{\nu_1}+\Delta m^2_{21}}+\sqrt{m^2_{\nu_1}+\Delta m^2_{31}}\,, \label{nu_mass1}\\
  \sum_{k=1}^3 m_{\nu_k}(\theta_2,\Delta m^2_2)&=&\sqrt{m^2_{\nu_2}-\Delta m^2_{21}}+m_{\nu_2}+\sqrt{m^2_{\nu_2}+\Delta m^2_{32}}\,, \label{nu_mass2}\\
 \sum_{k=1}^3 m_{\nu_k}(\theta_3,\Delta m^2_3)&=&m_{\nu_3}+\sqrt{m^2_{\nu_3}-\Delta m^2_{31}}+\sqrt{m^2_{\nu_3}-\Delta m^2_{32}}
 \label{nu_mass3}
\end{eqnarray}
where $m_{\nu_k}=\sqrt{\Delta m^2_k/\sin2\theta_k}\,(1+\sin2\theta_k)/2$.
Cosmological and astrophysical measurements provide powerful constraints on the sum of neutrino masses complementary to those from accelerators and reactors. 
Bounds on the sum of the three active neutrino masses can be summarized as
 \begin{eqnarray}
 0.06\,[{\rm eV}]\lesssim\sum_{i}m_{\nu_i}<\left\{
       \begin{array}{ll}
         0.340-0.715\,{\rm eV}\, & \hbox{(CMB PLANCK\,\cite{Moscibrodzka:2016ofe})} \\
         0.170\,{\rm eV}\, & \hbox{(CMB PLANCK+BAO\,\cite{Ade:2015xua})}\,,
       \end{array}
     \right.
  \label{nu_sum_b}
 \end{eqnarray}
a lower limit for the sum of the neutrino masses $\sum_{i=1}^{3} m_{\nu_i}\gtrsim0.06$ eV (for NO) and $\gtrsim0.103$ eV (for IO) could be provided by the neutrino oscillation measurements; upper limits\,\footnote{Massive neutrinos could leave distinct signatures on the CMB and large-scale structure (LSS) at different epochs of the Universe's evolution\,\cite{Abazajian:2008wr}. To a large extent, these signatures could be extracted from the available cosmological observations, from which the total neutrino mass could be constrained. } at $95\%$ CL are given in Ref.\,\cite{Moscibrodzka:2016ofe}.

\begin{figure}[t]
\begin{minipage}[h]{7.0cm}
\epsfig{figure=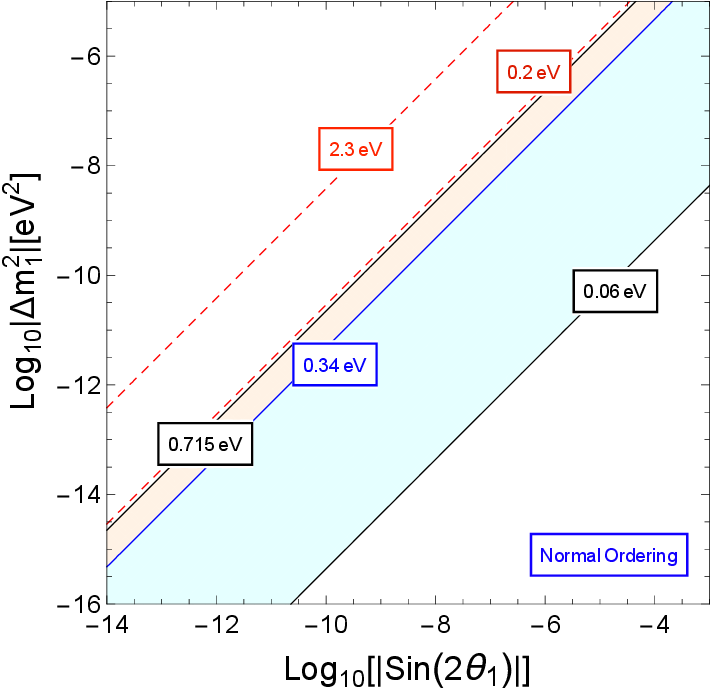,width=7.0cm,angle=0}
\end{minipage}
\hspace*{1.0cm}
\begin{minipage}[h]{7.0cm}
\epsfig{figure=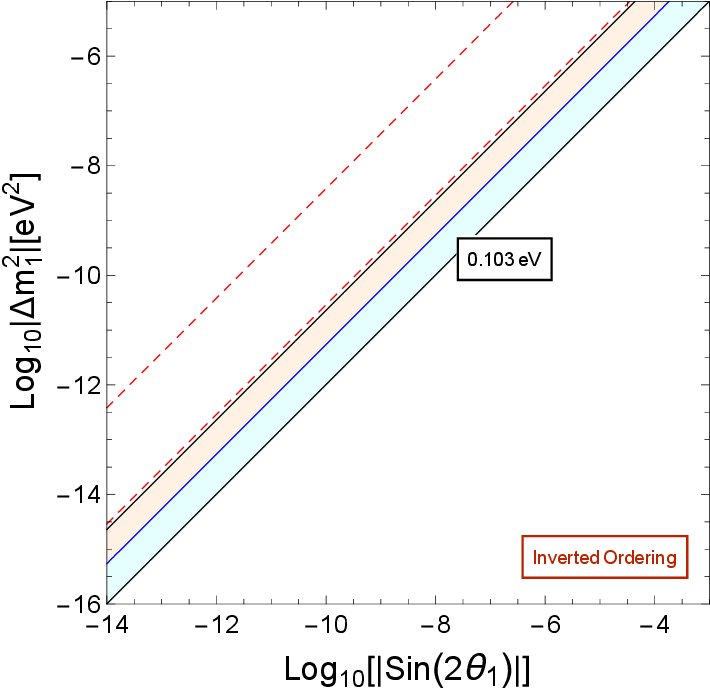,width=7.0cm,angle=0}
\end{minipage}\\
\vspace*{0.5cm}
\begin{minipage}[h]{7.0cm}
\epsfig{figure=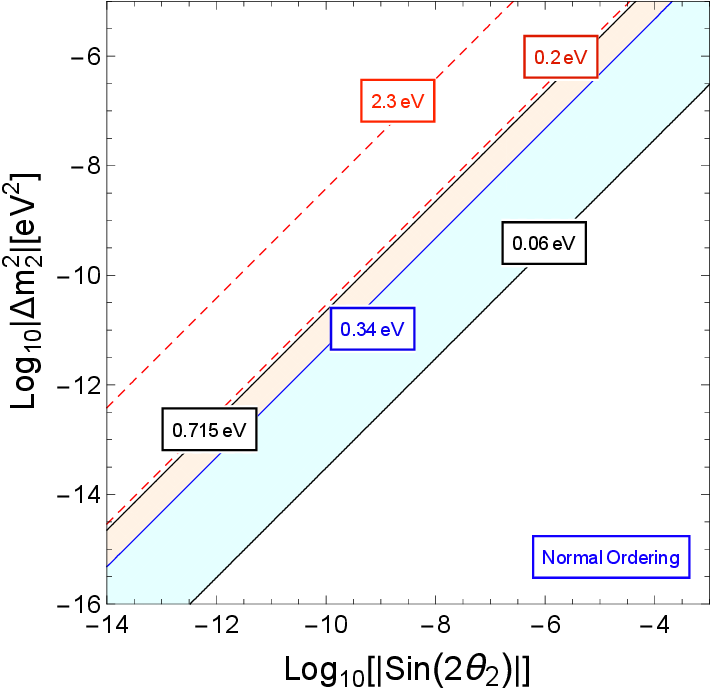,width=7.0cm,angle=0}
\end{minipage}
\hspace*{1.0cm}
\begin{minipage}[h]{7.0cm}
\epsfig{figure=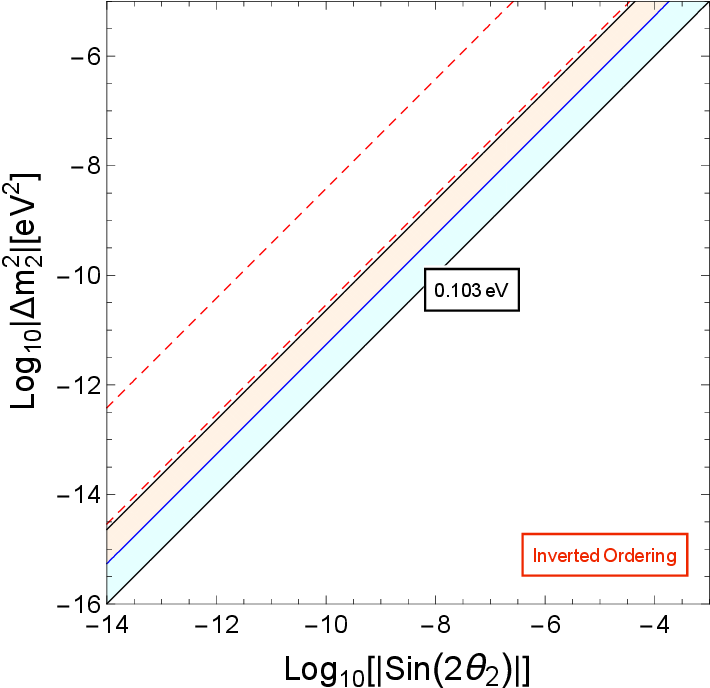,width=7.0cm,angle=0}
\end{minipage}
\caption{\label{Fig1} Contour plots in the parameter space ($\Delta{m}^2_{1}$, $\sin2\theta_1$: upper panel) and ($\Delta{m}^2_{2}$, $\sin2\theta_2$: lower panel) for fixed values of $\sum_{i=1,2,3} m_{\nu}$ (solid lines) and  $m_{\bar{\nu}_e}$ probed in tritium $\beta$ decay (dotted lines). The upper red-dotted line corresponds to the upper bound $m_{\bar{\nu}_e}<2.3$ eV\,\cite{Kraus:2004zw}, whereas the lower red-dotted line to a future sensitivity of $m_{\bar{\nu}_e}\lesssim0.20$\,\cite{Mertens:2015ila}. For $\sum_{i=1,2,3} m_{\nu}$, we take the values from Eq.(\ref{nu_sum_b}). And the best-fit values in Table-\ref{exp_nu} for the active neutrino oscillations are used.}
\end{figure}
The existence of massive neutrino at the eV scale can also be constrained by $\beta$-decay experiments\,\cite{Kraus:2012he} and by $0\nu\beta\beta$-decay experiments\,\cite{Giunti:2012tn}.
The two types of mass ordering, discussed above, should be compatible with the existing constraints on the absolute scale of neutrino masses $m_j$. The most sensitive experiments on the search of the effects of neutrino masses in $\beta$-decay use the tritium decay process $^3{\rm H}\rightarrow \,^3{\rm He}+e^-+\bar{\nu}_e$. Non-zero neutrino masses distort the measurable spectraum of the emitted electron. The most stringent  upper bounds on the $\bar{\nu}_e$ mass, $m_{\bar{\nu}_e}$, have been obtained from direct searches in the Mainz\,\cite{Kraus:2004zw} and Troitsk\,\cite{Aseev:2011dq} experiments at 95\% CL:
  \begin{eqnarray}
 m_{\bar{\nu}_e}=\Big(\sum_{k=1}^3|U_{ek}|^2m^2_{\nu_k}\Big)^{\frac{1}{2}}<\left\{
       \begin{array}{lll}
         2.30\,{\rm eV}\, ,& \hbox{Mainz\,\cite{Kraus:2004zw}} \\
         2.05\,{\rm eV}\, , & \hbox{Troitsk\,\cite{Aseev:2011dq}}\,.
       \end{array}
     \right.
   \label{tritum}
 \end{eqnarray}
The upcoming KATRIN experiment\,\cite{Mertens:2015ila} planned to reach sensitivity of $m_{\bar{\nu}_e}\sim0.20$ eV will probe the region of the QD spectrum in the model. Note that the bounds in Eq.\,(\ref{nu_sum_b}) coming from independent cosmological measurements are still tighter than the constraints coming from kinematic measurements of tritium $\beta$  decay. The $0\nu\beta\beta$-decay rate\,\cite{Aalseth:2004hb} effectively measures the absolute value of the $ee$-component of the effective neutrino mass matrix ${\cal M}_{\nu}$ in Eq.\,(\ref{nu_matr}) in the basis where the charged lepton mass matrix is real and diagonal. Since the two mass eigenstates of first and second generations in each pseudo-Dirac pair have opposite $CP$,  there appears only third generation in $\beta\beta0\nu$-decay rate. Using Eq.\,(\ref{nu_mass3}) one can easily see vanishing $\beta\beta0\nu$-decay rate, as shown in Ref.\,\cite{Ahn:2019jbm}. Hence if the $\beta\beta0\nu$-decay rate is measured in the near future the model would explicitly be excluded.

\begin{figure}[h]
\begin{minipage}[h]{7.0cm}
\epsfig{figure=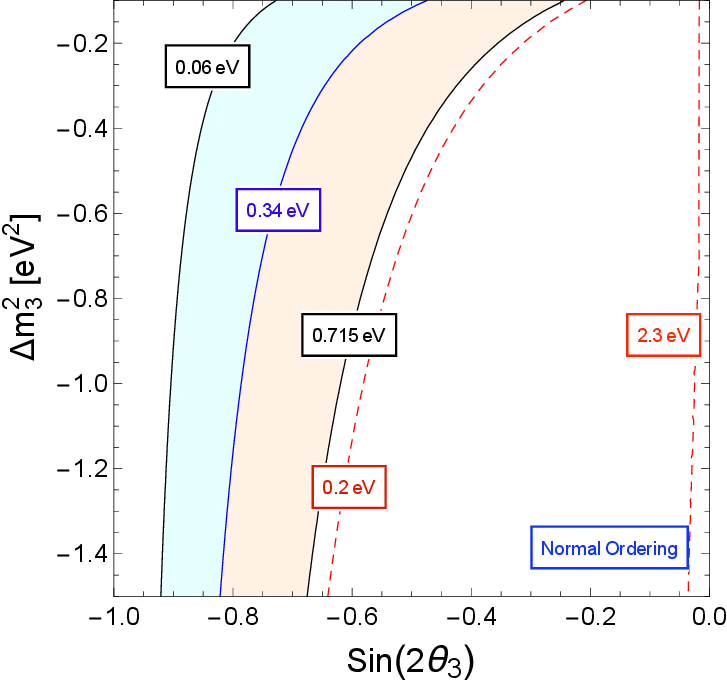,width=7.0cm,angle=0}
\end{minipage}
\hspace*{1.0cm}
\begin{minipage}[h]{7.0cm}
\epsfig{figure=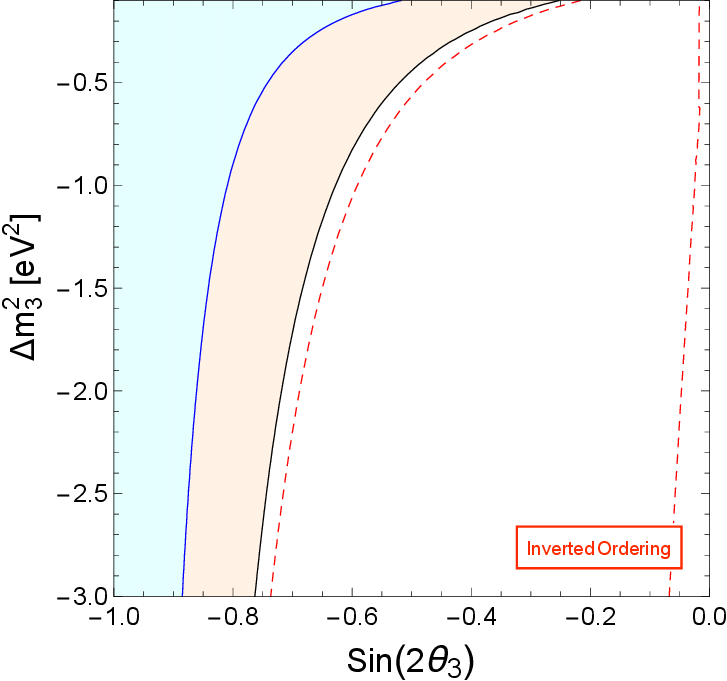,width=7.0cm,angle=0}
\end{minipage}
\caption{\label{Fig2} Same as Fig.\,\ref{Fig1} except for the parameter space ($\Delta{m}^2_{3}$, $\sin2\theta_3$).}
\end{figure}

In order to display new physical effects, we investigate the influence of $\Delta m^2_{k}$ and $\sin2\theta_k$ on the sum of active neutrino masses and the effective mass in $\beta$-decay. Plugging the experimental constraints of Table-\ref{exp_nu} into Eqs.\,(\ref{nu_mass1}-\ref{nu_mass3}) and Eq.\,(\ref{tritum}), allowed parameter spaces of $\Delta m^2_{k}$ and $\sin2\theta_k$ can be obtained in terms of the sum of active neutrino masses and the effective mass probed in tritium $\beta$-decay, respectively.
Contour plots in the parameter spaces ($\Delta m^2_{k}$, $\sin2\theta_k$) for fixed values of $\sum m_\nu$ (solid lines) and $m_{\bar{\nu}_e}$ probed in tritium $\beta$-decay (dotted lines) are presented for NO and IO in the upper panel of Fig.\,\ref{Fig1} for $\Delta m^2_{1}$ versus $\sin2\theta_1$ and in the lower panel of Fig.\,\ref{Fig1} for $\Delta m^2_{2}$ versus $\sin2\theta_2$, and Fig.\,\ref{Fig2} for $\Delta m^2_{3}$ versus $\sin2\theta_3$, where a lower limit for the sum of neutrino masses $\sum_{i=1}^{3} m_{\nu_i}\gtrsim0.06$ eV (for NO) and $\gtrsim0.103$ eV (for IO) and upper limits $0.340\sim0.715$ eV at $95\%$ CL\,\cite{Moscibrodzka:2016ofe}. In the plots we consider $10^{-16}<\Delta{m}^2_{1(2)}<10^{-5}\,{\rm eV}^2$ for Fig.\,\ref{Fig1} since they should be less than the measured $\Delta{m}^2_{\rm Sol}$, while for Fig.\,\ref{Fig2} only eV-mass scale of sterile neutrino since too heavy
neutrino is conflict with cosmology $\Delta{N}^{\rm eff}_\nu<0.2$ at $95\%$ CL\,\cite{Cyburt:2015mya, Hannestad:2012ky}.

\subsection{Interpretation of reactor neutrino at KamLAND}
As shown in Eq.\,(\ref{sol_kam}), the KamLAND $\bar{\nu}_e$ survival probability has reported a precise determination of $\Delta m^2_{\rm KL}=\Delta m^2_{21}$ and $\theta_{\rm KL}=\theta_{12}$ at $99.998\%$ C.L.\,\cite{Eguchi:2002dm, Araki:2004mb}. It has confirmed the LMA solution which can theoretically be explained via the MSW solar matter effects in neutrino oscillations\,\cite{Wolfenstein:1977ue, Mikheev:1986gs}.

Nuclear fission reactors are a powerful source of $\bar{\nu}_e$ with energies around a few MeV. Thus, the expected oscillation length is ${\cal O}(10\sim100)$ km, which is a reasonable distance relative to a reactor to place a detector and observe $\bar{\nu}_e$ disappearance. 
Assuming CPT (charge-conjugation, parity, and time-reversal) invariance, the oscillatory $\bar{\nu}_e$ signature observed by the KamLAND experiment\,\cite{kamland} can be reinterpreted in the baseline of $4\pi E/\Delta m^2_3\ll L\sim4\pi E/\{\Delta m^2_{21},\Delta S^2_{21}\}\ll 4\pi E/\Delta m^2_{2(1)}$ as 
\begin{eqnarray}
P_{\bar{\nu}_e\rightarrow\bar{\nu}_e}
&\approx&1-\cos^4\theta_{13}\sin^22\theta_{12}\sin^2\Big(\frac{\Delta m^2_{21}}{4E}L\Big)-\frac{1}{2}\big(\sin^22\theta_{13}+\sin^4\theta_{13}\cos^22\theta_3\big)\,,
\label{long03}
\end{eqnarray}
 which is CP conserved $P_{\bar{\nu}_e\rightarrow\bar{\nu}_e}=P_{\nu_e\rightarrow\nu_e}$ and does not include the earth matter effects\,\cite{PDG, Wolfenstein:1977ue, Mikheev:1986gs}, where $|\sin2\theta_2|\ll1$ and $\Delta m^2_{21}\approx\Delta S^2_{21}$ are used. The $\bar{\nu}_e$ disappearance probability depends on the parameters $\Delta m^2_{21}$ and $\theta_{12}$ as in the case of the $3\nu$SF, which derive the KamLAND $\bar{\nu}_e$ transitions, the dependence on $\theta_{13}$ and $\theta_3$ having. Due to $|\Delta m^2_3|\gg \Delta m^2_{21}\gg|\Delta m^2_{2}|$, at long-baselines (e.g. $\langle L\rangle\simeq180$ km) the terms involving $\sin^2\big(\frac{\Delta m^2_3}{4E}L\big)$ and $\sin^2\big(\frac{\Delta m^2_2}{4E}L\big)$ average out, such that Eq.\,(\ref{long03}) is sensitive to the parameters $\Delta m^2_{21}$ (or  $\Delta S^2_{21}$) and $\theta_{12}$, as the KamLAND $\nu_{\bar{e}}$ survival probability in the three neutrino standard form which is obtained in the limit $m_{s_i}\rightarrow m_{\nu_i}$ and $\theta_{i}\rightarrow0$ with $i=1,2,3$. Whereas it is negligible for the sensitivity of $\theta_3$ due to the tiny value of $\sin^4\theta_{13}\simeq5\times10^{-4}$. 
Thus, new effects due to the sterile neutrinos on KamLAND experiment can safely be negligible.
Hence the $\bar{\nu}_e$ survival probability of Eq.\,(\ref{long03}) is in good agreement with that of the three neutrino standard form $P^{3\nu{\rm SF}}_{\bar{\nu}_e\rightarrow\bar{\nu}_e}\simeq\sin^4\theta_{13}+\cos^4\theta_{13}\big(1-\sin^22\theta_{12}\sin^2\big(\frac{\Delta m^2_{21}}{4E}L\big)\big)$.

Within the $3\nu$SF\,\cite{Esteban:2018azc}, however, a clear discrepancy at $\lesssim2\sigma$ C.L. has been shown between the global solar neutrino data and  KamLAND reactor data regarding the value of $\Delta m^2$ with high accuracy in Eq.\,(\ref{sol_kam}), while the mixing angle is consistent. In the next section, we will claim that, if a new oscillation parameter $\Delta m^2_2$ in Eq.\,(\ref{assu}) is optimized by nature at the earth-sun distance $L_{es}=149.6\times10^6$ km and solar neutrino energies $0.1\lesssim E[{\rm MeV}]<19$, a new $\nu_e$ disappearance oscillation effect that only affect solar neutrinos could appear and give a solution to the solar neutrino tension.
 
\section{Solar neutrino Tension and its possible solution}
\noindent 
The low-energy $^8$B solar neutrinos (as low as $\sim3.5$ MeV) observed by SK\,\cite{Abe:2016nxk}, within the $3\nu$SF, translate into the mass-squared difference $\Delta m^2_{\odot}$ somewhat lower than $\Delta m^2_{\rm KL}$ measured by KamLAND but  the same mixing angle $\theta_{\odot}$ as $\theta_{\rm KL}$ measured by KamLAND, as shown in Eq.\,(\ref{sol_kam}). 
Moreover, it is well known that none of the $^8$B measurements performed by SNO\,\cite{Collaboration:2011nga}, SK\,\cite{Abe:2016nxk}, and Borexino\,\cite{Agostini:2017ixy} shows any evidence of the low energy spectrum turn-up expected in the standard MSW-LMA solution for the value of $\Delta m^2_{\rm KL}$ favored by KamLAND.
Hence the clear discrepancy at less than $2\sigma$\,\cite{Abe:2016nxk, Esteban:2018azc, Capozzi:2018ubv}  indicates that the current theoretical understanding of neutrino oscillation may not be complete. Some unknown physical phenomena that only affect solar neutrinos must be at play to explain such discrepancy.

 In this regard, we investigate why solar neutrinos at SNO\,\cite{Aharmim:2011vm}, SK\,\cite{Abe:2016nxk} and Borexino\,\cite{Agostini:2017ixy} appear to mix differently from reactor antineutrinos at KamLAND\,\cite{Eguchi:2002dm, Araki:2004mb} for theoretical understanding of the solar neutrino tension, by re-examining the MSW matter effects in our new framework of Eq.\,(\ref{lag}). 
\subsection{Solar matter effects in neutrino oscillations}
We discuss, first, how neutrinos produced in the Sun propagate towards the surface of the Sun and then to a detector on Earth, and study the relevant transition probabilities. To do this, we  construct an effective Hamiltonian for matter effects in the Sun by assuming that neutrinos propagating in matter interact coherently with the particles in the medium.
Neutrino propagation in matter is conveniently treated via a Schrodinger laboratory-frame time-evolution equation of the form $i\partial \psi_f/\partial x={\cal H}_m\,\psi_f$ where distance $x$ plays the role of time.  Here $\psi_f=(\nu_\alpha, S_\alpha)^T$ is a multi-component interaction state with correspondingly, the Hamiltonian in-matter ${\cal H}_m$ is a matrix in the interaction space $\psi_f$. The effective Hamiltonian in-matter ${\cal H}_m$ in the interaction basis has the form of $6\times6$ matrix
\begin{eqnarray}
 {\cal H}_m&=&\frac{1}{2E}\left[W^\ast_\nu{\left(\begin{array}{cc}
 m^{2}_{\nu_k}I_3 & 0_3  \\
 0_3 &  m^{2}_{s_k}I_3
 \end{array}\right)} W^T_\nu+{\left(\begin{array}{cc}
 A_{\alpha}I_3 & 0_3  \\
 0_3 &  0_{3}
 \end{array}\right)}\right]\quad\text{with}~A_\alpha=2E\, V_\alpha
\label{mat}
\end{eqnarray}
where $k=1,2,3$, $\alpha=e, \mu, \tau$, and $I_3$ $(0_3)$ stands for the $3\times3$ unit (null) matrix. Here the parameter $A_\alpha$ is a measure of the importance of matter effect with the matter-induced effective potential; $V_\alpha$ and $V_s=0$ are the potentials experienced by the active neutrinos and the sterile neutrinos, respectively, and $E$ is the neutrino energy. For anti-neutrino the Hamiltonian can be obtained by the substitution $V_\alpha\rightarrow-V_\alpha$ and $W_\nu\rightarrow W^\ast_\nu$. $\nu_e$'s have charged-current (CC) interactions with electrons and neutral-current (NC) interactions with nucleons $V_e=\sqrt{2}G_F(N_e-N_n/2)$, while $\nu_\mu$'s and $\nu_\tau$'s have only NC interactions $V_\mu=V_\tau=\sqrt{2}G_F(-N_n/2)$ whose equivalence is available at tree level in the weak interactions\,\cite{Botella:1986wy}, and any $S_\alpha$'s have no interactions, $V_s=0$, where $G_F$ is the Fermi constant, and $N_e$ ($N_n$) is the average electron (neutron) number per unit volume along the neutrino path. 

Consider, for example, electron neutrinos ($\nu_e$s) generated in the Sun by nuclear reactions. Assuming that there are no sterile neutrinos initially when the $\nu_e$s are produced by the weak interactions at the core of the Sun. The flavor state $\nu_e$ propagates as a mass eigenstate $\nu_{mi}$ in the medium and at any later time the eigenstate $\nu_{mi}$ has an ``active" and ``sterile" component. Hence, sterile neutrino gets generated as a vacuum-like state via the coherent oscillations with the assumption that no matter effects between active and sterile neutrinos occur during propagation. 

During neutrino propagation in matter, the mixing is defined with respect to eigenstates of the Hamiltonian in matter $n_{m}\equiv({\cal N}_{mi}, {\cal S}_{mi})^T$: ${\cal H}_m\,n_{m}={\cal H}_{mi}\,n_{m}$ where ${\cal H}_{mi}$ are the eigenvalues of ${\cal H}_m$. Thus a unitary mixing matrix $W_m$ in matter $(W_{m}W^\dag_m=W^\dag_{m}W_m=I)$ is defined as the matrix which connects the interaction states $\psi_f=(\nu_\alpha, S_\alpha)^T$ with the eigenstates of the Hamiltonian in matter $n_m$: 
\begin{eqnarray}
\psi_f={\left(\begin{array}{c}
 \nu_\alpha \\
 S_\alpha
 \end{array}\right)}=W_m{\left(\begin{array}{c}
 {\cal N}_{mi} \\
 {\cal S}_{mi}
 \end{array}\right)}\qquad\text{with}~W_m={\left(\begin{array}{cc}
 {\cal U}_LV_1 & i{\cal U}_LV_1  \\
 U_R{V}_2 &  -iU_R {V}_2 
 \end{array}\right)}V_\nu(\theta_k, \phi_k)\,,
\label{mat0}
\end{eqnarray}
where 
 ${\cal U}_L$ is a $3\times3$ transformation matrix of the form
\begin{eqnarray}
{\cal U}_L=\tilde{I}_1{\cal U}_{23}\,\tilde{I}_2{\cal U}_{13}\,\tilde{I}_3{\cal U}_{12}
\label{mat_u}
\end{eqnarray}
responsible for the mixing in matter for active neutrinos, which is a product of three matrices $\tilde{I}_k{\cal U}_{ij}\equiv\tilde{I}_k(\varphi^m_k)\,{\cal U}(\theta^m_{ij})$ rotation in corresponding planes by the set angles with the phase matrices $\tilde{I}_1={\rm diag}(1,e^{i\varphi^m_{1}}, e^{-i\varphi^m_{1}})$, $\tilde{I}_2={\rm diag}(e^{i\varphi^m_{2}}, 1, e^{-i\varphi^m_{2}})$, and $\tilde{I}_3={\rm diag}(e^{i\varphi^m_{3}}, e^{-i\varphi^m_{3}}, 1)$, and $U_R$ is given by the mixing matrix in vacuum since between the sterile neutrino themselves have no weak interactions in matter.

The mixing parameters in ${\cal U}_L$ and $V_\nu$ of Eq.\,(\ref{mat0}) appear in the final amplitudes of $\nu_\alpha\rightarrow\nu_\beta$ with $\alpha,\beta=e, \mu,\tau$, see Eq.\,(\ref{proba_matter}), when projecting the flavor states onto propagation basis states at the neutrino production and detection. Since the mixing parameters only in ${\cal U}_L$ (correspondingly instantaneous mass eigenstates) become functions of $V_\alpha$ for active neutrino propagation in a medium with varying density, it is expected that an active neutrino propagates  with both mixing angle between active neutrinos $\theta^m_{ij}$ and mixing angle between active and sterile neutrinos $\theta_k$.
Hence an intermediate instantaneous eigenstates $\xi_m\equiv(\nu_{mi}, S_{mi})^T$ can be defined with a transformation relation in matter
\begin{eqnarray}
 \psi_{f}={\left(\begin{array}{c}
 \nu_{\alpha} \\
 {S}_{\alpha}
 \end{array}\right)}= \tilde{W}_{m}{\left(\begin{array}{c}
 {\nu}_{mi} \\
 {S}_{mi}
 \end{array}\right)}\,\quad\text{with}~\tilde{W}_m={\left(\begin{array}{cc}
 {\cal U}_L & 0  \\
 0 &  U_R 
 \end{array}\right)}
\label{rela}
\end{eqnarray}
where a unitary mixing matrix $\tilde{W}_m$  transforms between the interaction eigenstates $\psi_f$ and the (instantaneous) mass eigenstates $\xi_m$. 
In turn, another unitary mixing matrix $X_m\equiv\tilde{W}^\dag_mW_m$ in matter can be defined as the matrix that connects the eigenstates $\xi_{m}$ with the eigenstates $n_{m}$:
\begin{eqnarray}
 \xi_{m}={\left(\begin{array}{c}
 {\nu}_{mi} \\
 {S}_{mi}
 \end{array}\right)}=X_m{\left(\begin{array}{c}
 {\cal N}_{mi} \\
 {\cal S}_{mi}
 \end{array}\right)}\quad\text{with}~X_m={\left(\begin{array}{cc}
 V_1 & iV_1  \\
 {V}_2 &  -i{V}_2 
 \end{array}\right)}V_\nu\,,
\label{mat1}
\end{eqnarray} 
in which the mixing matrix $X_m$ in matter between the active and sterile neutrinos becomes the one in vacuum. The sterile neutrinos get generated as vacuum-like states via coherent oscillations during propagation.

\subsubsection{Mixing in matter}
With the help of the mixing of Eq.\,(\ref{rela}), we can seclude the active neutrinos from the sterile neutrinos. 
In the interaction basis $\psi_f=(\nu_\alpha, S_\alpha)^T$, then, the effective Hamiltonian ${\cal H}_m$ of Eq.\,(\ref{mat}) in matter is modified to 
\begin{eqnarray}
 {\cal H}^a_m&=&\frac{1}{2E}\left[\tilde{W}^\ast_\nu{\left(\begin{array}{cc}
 m^{2}_{\nu_k}I_3 & 0_3  \\
 0_3 &  m^{2}_{s_k}I_3
 \end{array}\right)}\tilde{W}^T_\nu+{\left(\begin{array}{cc}
 A_{\alpha}I_3 & 0_3  \\
 0_3 &  0_{3}
 \end{array}\right)}\right]
\label{mat_a}
\end{eqnarray}
With respect to the eigenstates of the Hamiltonian in matter ${\cal H}^a_m$ of Eq.\,(\ref{mat_a}), $\xi_m$, {\it i.e.} ${\cal H}^a_m\xi_m={\cal H}^a_{mi}\xi_m$ where ${\cal H}^a_{mi}$ are the eigenvalues of ${\cal H}^a_{m}$, a relevant neutrino mixing in matter is given by the matrix $\tilde{W}_m$ in Eq.\,(\ref{rela}) that connects the interaction states $\psi_f$ with the eigenstates of the above Hamiltonian in matter $\xi_{m}$.

Then, the mixing of active neutrinos in matter can be defined with respect to the eigenstates of the active neutrino Hamiltonian in matter $\nu_{mi}$. 
In the basis of three neutrino flavors $\nu_\alpha=(\nu_e, \nu_\mu, \nu_\tau)^T$  the active neutrino Hamiltonian in matter $H_m$ has the form of $3\times3$ matrix
\begin{eqnarray}
 H_m&=&\frac{1}{2E}\left[U^\ast_L{\left(\begin{array}{ccc}
 0 & 0 & 0 \\
 0 &  \Delta{m}^{2}_{21} & 0\\
  0 &  0 & \Delta{m}^{2}_{31} 
 \end{array}\right)} U^T_L+{\left(\begin{array}{ccc}
 A_{e\mu} & 0 & 0  \\
 0 & 0& 0\\
  0 & 0 & 0
 \end{array}\right)}\right]\,,
\label{mat_H}
\end{eqnarray}
where irrelevant terms of the Hamiltonian proportional to the unit matrix are omitted. 
Considering the matter effects in the Sun, which are of special relevance for solar neutrinos. As $\nu_e$s produced at the core of the Sun move outward, $A_{e\mu}$ will decrease as the density decreases, and the neutrinos will eventually go through the resonance regions, and proceed out into the vacuum.
The electron (neutron) number density $N_{e}$ $(N_n)$ of solar matter measured in number per cm$^3$ is a monotonically decreasing function of the distance $R$ from the center of the Sun along the neutrino path, and they can be analytically approximated as $\log(N_{e}/N_A)=2.36-4.52R/R_{\odot}-0.33\,e^{-(R/0.075R_{\odot})^{1.1}}$ and $\log(N_n/N_A)=1.72-4.80R/R_\odot$ with the solar radius $R_{\odot}=6.96\times10^5$ km and Avogadro's number $N_A$\,\cite{Bahcall:2000nu, Bahcall:2004fg, Bahcall:2006}, where the logarithm is base 10 and $N_e(0)\simeq103\,N_A/{\rm cm}^3$ ($N_n(0)\simeq50\,N_A/{\rm cm}^3$) is the electron (neutron) number density at the point of $\nu_e$ production in the Sun\,\cite{Bahcall:2006}. 
Then the difference of the potentials for $\nu_e$ and $\nu_{\mu,\tau}$, {\it i.e.} $A_{e\mu}\equiv A_e-A_\mu=2E(V_e-V_\mu)$, due to the charged current scattering of $\nu_e$ on electrons ($\nu_ee\rightarrow\nu_ee$)\,\cite{Wolfenstein:1977ue}, reads $A_{e\mu}=2\sqrt{2}\,E\,G_FN_e(R)$:
\begin{eqnarray}
A_{e\mu}(R,E)\simeq1.57\times10^{-5}\Big(\frac{N_{e}(R)}{103\,N_A{\rm cm}^{-3}}\Big) \Big(\frac{E}{1{\rm MeV}}\Big){\rm eV}^2\,.
\label{Ae_mat}
\end{eqnarray}
After crossing the Sun, the make-up of the neutrino state existing the Sun will depend on the relative size of $A_{e\mu}$ versus $A^r_{ij}$ (at neutrino-state resonance point) whose parameters are determined by nature. 
Similar to Eq.\,(\ref{mat_u}) in matter, the mixing matrix in vacuum $U_L$ in Eq.\,(\ref{mat_H}) is a general form of mixing matrix for three neutrino flavors, instead of the standard form of Eq.\,(\ref{PMNS}),
\begin{eqnarray}
 U_L=I_1U_{23}\,I_2U_{13}\,I_3U_{12}
\label{gen_U}
\end{eqnarray}
as a product of three matrices $I_kU_{ij}\equiv I_k(\varphi_k)U(\theta_{ij})$ of rotation in the space of neutrino mass eigenstates by the set angles with the phase matrices $I_1={\rm diag}(1,e^{i\varphi_{1}}, e^{-i\varphi_{1}})$, $I_2={\rm diag}(e^{i\varphi_{2}}, 1, e^{-i\varphi_{2}})$, and $I_3={\rm diag}(e^{i\varphi_{3}}, e^{-i\varphi_{3}}, 1)$.

The mixing in matter for active neutrinos is defined with respect to the eigenstates of the active neutrino Hamiltonian $\nu_{mi}={\cal U}^\dag_L\nu_\alpha$: $H_m\,\nu_{mi}={H}_{mi}\,\nu_{mi}$ where ${H}_{mi}$ are  the eigenvalues of ${H}_m$.
The effective matter mass eigenstates $|\nu_{mi}\rangle$ with effective masses $M_{\nu_i}$ satisfy the eigenequation $H_m|\nu_{mi}\rangle=\frac{M^2_{\nu_i}}{2E}|\nu_{mi}\rangle$.
 The Hamiltonian in Eq.\,(\ref{mat_H}) can be diagonalized performing several consecutive rotations with Eq.\,(\ref{mat_u})
\begin{eqnarray}
 {H}_{mi}={\cal U}^\dag_LH_m\,{\cal U}_L\,.
\label{mat_eigen}
\end{eqnarray}
Due to the matter effect of the form of ${\rm diag}(A_{e\mu}, 0, 0)$ in Eq.\,(\ref{mat_H}), the 2-3 rotations do not depend on the effect of medium  on neutrinos, leading to $\theta^m_{23}=\theta_{23}$ at leading order and $\varphi^m_1=\varphi_1$. Furthermore\,\footnote{In the charged-lepton basis, the Dirac CP phase $\delta_{CP}$ in the standard form of Eq.\,(\ref{PMNS}) is equivalent to $\delta_{CP}=\varphi_1-\varphi_2$ in the general form of Eq.\,(\ref{gen_U}) due to $\varphi_3=0$.}, the diagonalization of Eq.\,(\ref{mat_eigen}) requires $\varphi^m_2=\varphi_2$ and $\varphi^m_3=\varphi_3=0$. Then, due to the perturbative primary approximation one can obtain effective mixing angles in matter 
\begin{eqnarray}
 \tan2\theta^m_{13}&=&\frac{\sin2\theta_{13}}{\cos2\theta_{13}+\frac{A_{e\mu}}{\Delta{m}^2_{21}\sin^2\theta_{12}-\Delta{m}^2_{31}}}\,,\label{mat_13}\\
 \tan2\theta^m_{12}&=&\frac{\sin2\theta_{12}}{\frac{1}{\cos(\theta^m_{13}-\theta_{13})}\big(\cos2\theta_{12}-\frac{A_{e\mu}}{\Delta{m}^2_{21}}\cos^2\theta^m_{13}\big)+\frac{\sin^2(\theta^m_{13}-\theta_{13})}{\cos(\theta^m_{13}-\theta_{13})}\big(\sin^2\theta_{12}-\frac{\Delta{m}^2_{31}}{\Delta{m}^2_{21}}\big)}
\label{mat_12}
\end{eqnarray}
where higher order corrections are proportional to $\sin(\theta_{13}-\theta^m_{13})$, and mass-squared eigenvalues in matter
\begin{eqnarray}
 M^2_{\nu_1}&=& A_{e\mu}\cos^2\theta^m_{13}\cos^2\theta^m_{12}+\Delta\tilde{m}^2_{31}\sin^2(\theta^m_{13}-\theta_{13})\cos^2\theta^m_{12}+\Delta{m}^2_{21}\Big\{\cos^2\theta_{12}\sin^2\theta^m_{12}\nonumber\\
 &&~+\sin^2\theta_{12}\cos^2(\theta^m_{13}-\theta_{13})\cos^2\theta^m_{12}-\frac{1}{2}\sin2\theta_{12}\sin2\theta^m_{12}\cos(\theta_{13}-\theta^m_{13})\Big\}+{m}^2_{\nu_1}\,,\nonumber\\
 M^2_{\nu_2}&=&A_{e\mu}\cos^2\theta^m_{13}\sin^2\theta^m_{12}+\Delta{m}^2_{31}\sin^2(\theta^m_{13}-\theta_{13})\sin^2\theta^m_{12}+\Delta{m}^2_{21}\Big\{\cos^2\theta_{12}\cos^2\theta^m_{12}\nonumber\\
 &&~+\sin^2\theta_{12}\cos^2(\theta^m_{13}-\theta_{13})\sin^2\theta^m_{12}+\frac{1}{2}\sin2\theta_{12}\sin2\theta^m_{12}\cos(\theta_{13}-\theta^m_{13})\Big\}+ {m}^2_{\nu_1}\,,\nonumber\\
 M^2_{\nu_3}&=&A_{e\mu}\sin^2\theta^m_{13}+\Delta{m}^2_{21}\sin^2\theta_{12}\sin^2(\theta_{13}-\theta^m_{13})+\Delta{m}^2_{31}\cos^2(\theta_{13}-\theta^m_{13})+{m}^2_{\nu_1}\,,
\label{mat_eis}
\end{eqnarray}
where higher order corrections proportional to $\sin^2(\theta_{13}-\theta^m_{13})$ are neglected.

In fact, through a 1-3 rotation in Eq.\,(\ref{mat_eigen}) the vanished 2-3 and 1-3 elements of matrix ${H}_{mi}$ are corrected as $\frac{1}{2}\Delta\tilde{m}^2_{21}\sin2\theta_{12}\sin(\theta^m_{13}-\theta_{13})\cos\theta^m_{12}$ and $\frac{1}{2}\Delta\tilde{m}^2_{21}\sin2\theta_{12}\sin(\theta^m_{13}-\theta_{13})\sin\theta^m_{12}$, respectively, reflecting that both corrections are proportional to $\sin(\theta^m_{13}-\theta_{13})$ and whose corrections should vanish by doing additional 2-3 and 1-3 rotations, but they can be safely neglected in the Sun, see below.
From Eqs.\,(\ref{mat_13}) and (\ref{mat_12}), as usual, we define resonant values
\begin{eqnarray}
 A^r_{13}\equiv\cos2\theta_{13}(\Delta{m}^2_{31}-\sin^2\theta_{12}\Delta{m}^2_{21})\,,\quad A^r_{12}\equiv\Delta{m}^2_{21}\frac{\cos2\theta_{12}}{\cos^2\theta^m_{13}}\,.
\label{reso_r}
\end{eqnarray}
Using Table-\ref{exp_nu} and Eq.\,(\ref{Ae_mat}) for given solar neutrino energies $0.1\lesssim E[{\rm MeV}]<19$, from Eq.\,(\ref{mat_13}) we obtain 
\begin{eqnarray}
 \theta^m_{13}\simeq\theta_{13}
\label{th13}
\end{eqnarray}
with good accuracy, since the condition of $A^r_{13}\gg A_{e\mu}(R, E)$ is satisfied in all ranges of solar neutrino energies. 
Thus the corrections to the $\theta_{23}$ and $\theta^m_{13}$ can be safely neglected in solar matter.
Then each equation in Eq.\,(\ref{reso_r}) can be approximated with good accuracy as $A^r_{13}\simeq\Delta m^2_{31}\cos2\theta_{13}$ and $A^r_{12}\simeq\Delta{m}^2_{21}\cos2\theta_{12}/\cos^2\theta_{13}$.

Therefore, in solar matter a matter mixing angle $\theta^m_{12}$ is only effective, and the mixing matrix Eq.\,(\ref{mat_u}) becomes
\begin{eqnarray}
 {\cal U}_L\simeq I_1U_{23}(\theta_{23})\,I_2U_{13}(\theta_{13})\,I_3{\cal U}_{12}(\theta^m_{12})\,,
\label{new_mat12}
\end{eqnarray}
and the mass-squared eigenvalues in Eq.\,(\ref{mat_eis}) become
\begin{eqnarray}
 M^2_{\nu_1}&\simeq& {m}^2_{\nu_1}+A_{e\mu}\cos^2\theta_{13}\cos^2\theta^m_{12}+\Delta{m}^2_{21}\sin^2(\theta_{12}-\theta^m_{12})\,,\nonumber\\
 M^2_{\nu_2}&\simeq&{m}^2_{\nu_1}+A_{e\mu}\cos^2\theta_{13}\sin^2\theta^m_{12}+\Delta{m}^2_{21}\cos^2(\theta_{12}-\theta^m_{12})\,,\nonumber\\
 M^2_{\nu_3}&\simeq&{m}^2_{\nu_1}+ A_{e\mu}\sin^2\theta_{13}+\Delta{m}^2_{31}\,.
\label{mat_eis1}
\end{eqnarray}
In vacuum limit {\it i.e.} $\theta^m_{ij}\rightarrow\theta_{ij}$ with $A_{e\mu}\rightarrow0$, from Eq.\,(\ref{mat_eis}) we clearly see that the mass-squared eigenvalues $M^2_{\nu_i}$ go to $m^2_{\nu_i}$.
From Eqs.\,(\ref{mat_12}) and (\ref{mat_eis}), then, the mixing angle $\theta^m_{12}$ and the mass-squared difference $\Delta M^2_{21}=M^2_{\nu_2}-M^2_{\nu_1}$ are well approximated to
\begin{eqnarray}
 \tan2\theta^m_{12}&\simeq&\frac{\sin2\theta_{12}}{\cos2\theta_{12}-\frac{A_{e\mu}\cos^2\theta_{13}}{\Delta{m}^2_{21}}}\,,
\label{mat1_12}\\
\Delta M^2_{21}&\simeq&\Delta{m}^{2}_{21}\Big[\Big(\cos2\theta_{12}-\frac{A_{e\mu}\cos^2\theta_{13}}{\Delta{m}^2_{21}}\Big)^2+\sin^22\theta_{12}\Big]^{\frac{1}{2}}
\label{mat1_120}\,.
\end{eqnarray}

For a medium with varying density the eigenstates $\nu_{mi}$ are no longer eigenstates of the Hamiltonian of Eq.\,(\ref{mat_H}). Indeed, since ${\cal U}_L$ is $x$ dependent, the neutrino propagation equation is written as
\begin{eqnarray}
 i\frac{d\,\nu_{mi}}{dx}=\Big({H}_{mi}-i\,{\cal U}_L^\dag\frac{d\,{\cal U}_L}{dx}\Big)\nu_{mi}\,.
\label{nuprop2}
\end{eqnarray}
Considering Eq.\,(\ref{new_mat12}) reduces the above neutrino propagation equation to the two-neutrino state problem in solar matter
\begin{eqnarray}
 i\frac{d}{dx}{\left(\begin{array}{c}
 {\nu}_{m1} \\
 {\nu}_{m2}
 \end{array}\right)}={\left(\begin{array}{cc}
 -\Delta M^2_{21}/4E & -i\,d\theta^m_{12}/dx  \\
 i\,d\theta^m_{12}/dx &  \Delta M^2_{21}/4E 
 \end{array}\right)}{\left(\begin{array}{c}
 {\nu}_{m1} \\
 {\nu}_{m2}
 \end{array}\right)}\,,
\label{nuprop21}
\end{eqnarray}
where the eigenstates are redefined as $\nu_{mi}\rightarrow e^{i(M^2_{\nu_1}+M^2_{\nu_2})/4E}\nu_{mi}$ with $i=1,2$ and
\begin{eqnarray}
 \frac{d\theta^m_{12}}{dx}=\frac{1}{2}\frac{\Delta\,\sin2\theta_{12}}{(\Delta\,\cos2\theta_{12}-A_{e\mu})^2+(\Delta\,\sin2\theta_{12})^2}\frac{dA_{e\mu}}{dx}\quad\text{with}~\Delta\equiv\frac{\Delta{M}^2_{21}}{\cos^2\theta_{13}}\,.
\label{nuprop22}
\end{eqnarray}
If the density is slowly changing, on a distance scale of roughly the wavelength in matter, the off-diagonal term $d\theta^m_{12}/dx$ can be negligible.
Here we assume the adiabatic condition $|d\theta^m_{12}/dx|\ll|\Delta{M}^2_{21}|/4E$, then the states $\nu_{mi}$ become the eigenstates of the Hamiltonian of Eq.\,(\ref{mat_H}). 

Next, to find a mixing between the active and sterile neutrinos we perform a basis rotation
\begin{eqnarray}
\psi_f= {\left(\begin{array}{c}
 {\nu}_{\alpha} \\
 {S}_{\alpha}
 \end{array}\right)}\rightarrow {\left(\begin{array}{c}
 {\nu}_{i} \\
 {S}_{i}
 \end{array}\right)}=\tilde{W}_\nu{\left(\begin{array}{c}
 {\nu}_{\alpha} \\
 {S}_{\alpha}
 \end{array}\right)}\quad\text{with}~\tilde{W}_\nu={\left(\begin{array}{cc}
 U_L & 0_3  \\
 0_3 &  U_R
 \end{array}\right)}\,.
\label{basis_ch}
\end{eqnarray}
 Then the associated Hamiltonian for the active to sterile neutrinos in vacuum is given in the mass eigenstates $\xi=(\nu_i, S_i)^T$ by
\begin{eqnarray}
 {\cal H}^{as}&=&\frac{1}{2E}X^\ast_\nu{\left(\begin{array}{cc}
 m^{2}_{\nu_k}I_3 & 0_3  \\
 0_3 &  m^{2}_{s_k}I_3
 \end{array}\right)} X^T_\nu\,,
\label{hamv}
\end{eqnarray}
where $X_\nu=\tilde{W}^\dag_\nu W_\nu$.  
In the propagation basis $\xi_m=({\nu}_{mi}, S_{mi})^T$ of Eq.\,(\ref{mat1}), the associated Hamiltonian in matter is given with the replacement of $m^2_{\nu_k}$ by the effective mass $M^{2}_{\nu_k}$ of Eq.\,(\ref{mat_eis1}) as
\begin{eqnarray}
 {\cal H}^{as}_m&=&\frac{1}{2E}X^\ast_\nu{\left(\begin{array}{cc}
 M^{2}_{\nu_k}I_3 & 0_3  \\
 0_3 &  m^{2}_{s_k}I_3
 \end{array}\right)} X^T_\nu\,,
\label{mat11}
\end{eqnarray}
where $X_\nu$ is equivalent to $X_m$ in Eq.\,(\ref{mat1}). Eq.\,(\ref{mat11}) shows clearly that, in the propagation basis $\xi_m$, for example, an effective mass of an electron flavor neutrino is not induced by a background electron medium.
Hence, it is assumed that  there is no matter effect between active and sterile neutrinos during propagation in the Sun, as mentioned before.
The mixing in matter $X_m$ in Eq.\,(\ref{mat1}) can be reduced to two-neutrino state problem between the active neutrino and the sterile neutrino, which can be defined with respect to the eigenstates of the effective Hamiltonian ${\cal H}^{as}_m$ for the active to sterile neutrinos $n_{m}=({\cal N}_{mi}, {\cal S}_{mi})^T$: ${\cal H}^{as}_m\,n_{m}={\cal H}^{as}_{mi}\,n_{m}$ where ${\cal H}^{as}_{mi}$ are the eigenvalues of ${\cal H}^{as}_{m}$. Then the effective in-matter mass eigenstates $|n_{m}\rangle$ with the masses ${M}_{\nu_k}$ and ${m}_{s_k}$ satisfy the eigenequation 
\begin{eqnarray}
{\cal H}^{as}_m|n_{m}\rangle=\frac{1}{2E}{\left(\begin{array}{cc}
 M^2_{\nu_k} & 0_3  \\
 0_3 &  m^2_{s_k}
 \end{array}\right)}|n_{m}\rangle\,.
\label{egen}
\end{eqnarray}
Note that the eigenvalues in the eigenequation ${\cal H}^{as}_m|n_m\rangle={\cal H}^{as}_{mi}|n_m\rangle$ do not have matter effects and behave like those in vacuum.

\subsubsection{Three-active-neutrino Oscillation  Probabilities in Matter}
We are interested in the flavor transition between the active neutrinos $\nu_e$, $\nu_\mu$, $\nu_\tau$.
The transition probability in matter between the massive neutrinos that a neutrino eigenstate $\nu_\alpha$ becomes an eigenstate $\nu_\beta$ follows from the time evolution of mass eigenstates as
\begin{eqnarray}
P^m_{\nu_\alpha\rightarrow\nu_\beta}(\theta_k, \theta^m_{ij}, \theta_{ij}, L, E)&=&\Big|(W_m\,e^{-i\frac{\tilde{\cal M}^2_\nu}{2E}L}W^\dag_\nu)_{\alpha\beta}\Big|^2\,,
\label{proba_matter}
\end{eqnarray}
where $\alpha, \beta$ denote either $e$, $\mu$, or $\tau$, and $\tilde{\cal M}^2_\nu={\rm diag}(M^2_{\nu_1}, M^2_{\nu_2}, M^2_{\nu_3}, {m}^2_{s_1}, {m}^2_{s_2}, {m}^2_{s_3})$ in Eq.\,(\ref{egen}). We have assumed adiabatic case in Eq.\,(\ref{nuprop21}) so that $n_{m}=({\cal N}_{mi}, {\cal S}_{mi})^T$ propagate from production to the surface of matter and the mass eigenstates in matter do not mix. 
From Eq.\,(\ref{proba_matter}) the flavor transition probability between the active neutrinos $\nu_{e,\mu,\tau}$ can be explicitly expressed in terms of the oscillation parameters $\theta$, $\theta^m$, $L$, $E$, and $\Delta m^2$ as
\begin{eqnarray}
P^m_{\nu_\alpha\rightarrow\nu_\beta}&=&\sum^3_{k=1}|{\cal U}_{\alpha k}|^2|{U}_{\beta k}|^2\Big\{\cos^2\Big(\frac{\Delta{m}^2_k}{4E}L\Big)+\sin^22\theta_k\sin^2\Big(\frac{\Delta{m}^2_k}{4E}L\Big)\Big\}\nonumber\\
&+&\sum_{k>j}{\rm Re}\big[{U}^\ast_{\beta k}{U}_{\beta j}{\cal U}^\ast_{\alpha j}{\cal U}_{\alpha k}\big]\big(2-\sin^2\tilde{\Delta}_{kj}\big)
+\frac{1}{2}\sum_{k>j}{\rm Im}\big[{U}^\ast_{\beta k}{U}_{\beta j}{\cal U}^\ast_{\alpha j}{\cal U}_{\alpha k}\big]\sin\tilde{\Delta}_{kj}\,,
\label{osc01_matter}
\end{eqnarray}
where $\sin^2\tilde{\Delta}_{kj}$ and $\sin\tilde{\Delta}_{kj}$ are given by Eqs.\,(\ref{osc02}) and (\ref{osc03}), respectively,
except for the replacement of $\Delta m^2_{kj}$ by $\Delta M^2_{kj}=M^2_{\nu_k}-M^2_{\nu_j}$, and the $3\times3$ mixing components $U_{\alpha i}\equiv U_{\alpha i}(L)$ in vacuum at neutrino detection point $L$ and ${\cal U}_{\alpha i}\equiv {\cal U}_{\alpha i}(x_0)$ in matter at neutrino production point $x_0$. 
As expected, in the limit of $\theta^m_{ij}\rightarrow\theta_{ij}$ ($i\neq j=1,2,3$) with $A_{e\mu}\rightarrow0$, the transition probability in matter of Eq.\,(\ref{osc01_matter}) is recovered to the form of Eq.\,(\ref{osc01}) in vacuum; moreover, with the limit of $\Delta{m}_k\rightarrow0$ and $\theta_k\rightarrow0$ it is recovered to the $3\nu$SF, {\it i.e.}
the standard form of that for three active neutrinos, in vacuum\,\cite{PDG}.

\begin{figure}[h]
\hspace*{-1.0cm}
\begin{minipage}[h]{8.5cm}
\epsfig{figure=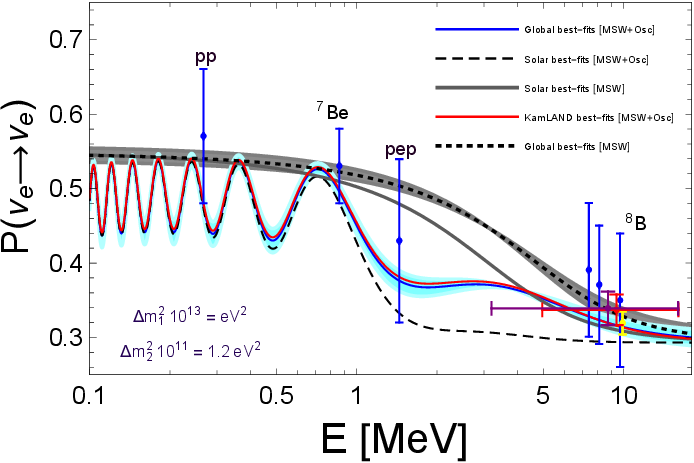,width=8.5cm,angle=0}
\end{minipage}
\hspace*{0.1cm}
\begin{minipage}[h]{8.5cm}
\epsfig{figure=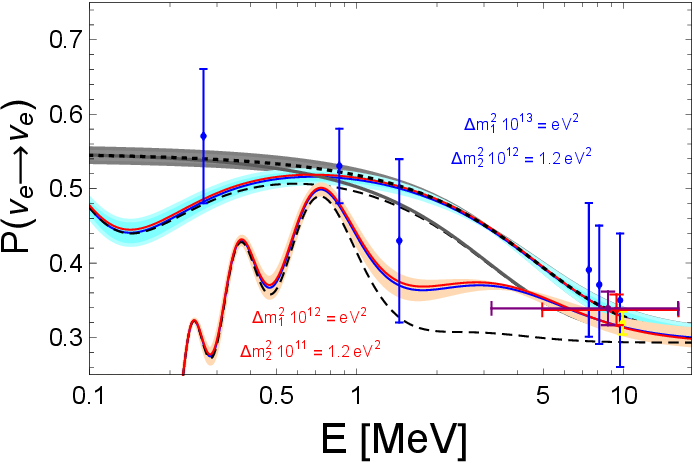,width=8.5cm,angle=0}
\end{minipage}
\caption{\label{Fig4} Solar $\nu_e$ survival probability as a function of neutrino energy, where $N_e\simeq85.5\,N_A{\rm cm}^{-3}$ in Eq.\,(\ref{Ae_mat}) at around the center of the Sun is used. The points
represent, from left to right, the Borexino $pp$, $^7$Be, $pep$, and $^8$B data (blue points) given in\,\cite{napp} and the SNO+SK
$^8$B data (yellow point) given in\,\cite{PDG}. Also shown are the SNO+SK data (red point) and SNO LETA + Borexino (purple point) of the $^8$B flux\,\cite{Collaboration:2011nga}. The error bars represent the $1\sigma$ experimental + theoretical uncertainties. The gray shaded band corresponds to the prediction of the MSW-LMA solution of the $3\nu$SF using the $1\sigma$ parameter values given in Table-\ref{exp_nu} and the black dotted line for the best-fits, while the gray line corresponds to the solar best-fits of Eq.\,(\ref{sol_kam}).
In left plot for new predictions of the MSW+ Oscillation effects for $\Delta m^2_1=10^{-13}\,{\rm eV}^2$ and $\Delta m^2_2=1.2\times10^{-11}\,{\rm eV}^2$, the cyan band corresponds to the $1\sigma$ parameter values in Table-\ref{exp_nu} and the blue solid line for the global best-fits, while the red solid line corresponds to the KamLAND best-fits of Eq.\,(\ref{sol_kam}) and the black dashed line for the solar best-fits of Eq.\,(\ref{sol_kam}). In right plot, the cyan band (blue solid, red solid, and black dashed lines) correspond to $\Delta m^2_1=10^{-13}\,{\rm eV}^2$ and $\Delta m^2_2=1.2\times10^{-12}\,{\rm eV}^2$, while the orange band (blue solid, red solid, and black dashed lines) for $\Delta m^2_1=10^{-12}\,{\rm eV}^2$ and $\Delta m^2_2=1.2\times10^{-11}\,{\rm eV}^2$.}
\end{figure}

Now we have a new theoretical framework for neutrino oscillation that allows to compute the probability of neutrinos born with one flavor in matter tuning into a different flavor.
Let us discuss how our scenario can describe the solar neutrino tension with the MSW-LMA solar neutrino oscillations. Considering baselines $4\pi E/\{\Delta{m}^2_3, \Delta{Q}^2_{3\ell}, \Delta{S}^2_{3\ell}\}\ll 4\pi E/\{\Delta M^2_{21}, \Delta{Q}^2_{21}, \Delta{Q}^2_{12}, \Delta{S}^2_{21}\}\ll L$ (where $\ell=1,2$) and averaging out the associated oscillating phases in the propagation between the Sun and the Earth for given solar neutrino energies, the solar $\nu_e$ transition at the exposed surface of the Earth can be described by
\begin{eqnarray}
P^m_{\nu_e\rightarrow\nu_e}&\simeq&\cos^2\theta_{13}\cos^2\theta^m_{13}\Big\{\cos^2\theta^m_{12}\cos^2\theta_{12}\cos^2\Big(\frac{\Delta{m}^2_1}{4E}L\Big)\nonumber\\
&&\qquad+ \sin^2\theta^m_{12}\sin^2\theta_{12}\cos^2\Big(\frac{\Delta{m}^2_2}{4E}L\Big)\Big\}+\frac{1}{2}\sin^2\theta^m_{13}\sin^2\theta_{13}(1+\sin^22\theta_3)\,,
\label{m_Ps}
\end{eqnarray}
where the sensitivity of $\theta_3$ is negligible due to the tiny value of $\sin^4\theta_{13}\simeq5\times10^{-4}$ and $\theta^m_{13}\simeq\theta_{13}$ of Eq.\,(\ref{th13}).
The above $\nu_e$ transition probability is different from that of the conventional $3\nu$SF, $P^{m3\nu{\rm SF}}_{\nu_e\rightarrow\nu_e}\simeq\cos^2\theta_{13}\cos^2\theta^m_{13}\big(\cos^2\theta^m_{12}\cos^2\theta_{12}+\sin^2\theta^m_{12}\sin^2\theta_{12}\big)+\sin^2\theta_{13}\sin^2\theta^m_{13}$\,\cite{PDG}, in that it contains oscillatory terms of $\Delta{m}^2_{1(2)}$.
We refer to the matter effect that causes a neutrino flavor change via both the effects of sinusoidal oscillation and the MSW matter effect as a ``composite matter effect". Contrary to the conventional MSW matter effect that causes a change of electron neutrino but without sinusoidal oscillating terms\,\cite{Wolfenstein:1977ue, Mikheev:1986gs}, the so-called composite matter effect causes an electron neutrino change via the effects of sinusoidal oscillation induced by the oscillatory terms of $\Delta{m}^2_{1(2)}$, as well as the MSW matter effect, as shown in Eq.\,(\ref{m_Ps}).
In the $\nu_e$ transition probability of Eq.\,(\ref{m_Ps}) from the composite matter effect to vacuum oscillations, we will show that the value of $\Delta m^2_{\rm KL}$ measured in KamLAND\,\cite{Eguchi:2002dm, Araki:2004mb} can be compatible with that measured in SNO\,\cite{Aharmim:2011vm}, SK\,\cite{Abe:2016nxk}, and Borexino\,\cite{Agostini:2017ixy} at higher energies ($>3$ MeV) through new oscillation effects induced by the terms containing $\Delta{m}^2_{1(2)}$:
\begin{itemize}
\item 
If the matter parameter $A_{e\mu}$ at $\nu_e$ production point is much larger than the resonant density, {\it i.e.} $A_{e\mu}(x_0, E)\gg A^r_{12}$ in Eq.\,(\ref{mat1_12}), the matter mixing angle $\theta^m_{12}$ of Eq.\,(\ref{new_mat12}) goes to $\pi/2$ for $\Delta{m}_{21}>0$ and $\cos2\theta_{12}>0$ $(\theta_{12}<\pi/4)$. 
Then, the solar $\nu_e$ transition can be described\,\footnote{In this case, the oscillatory term containing $\Delta m^2_1$ is disappeared in the limit of $\theta^m_{12}\rightarrow\pi/2$.} as
\begin{eqnarray}
P^m_{\nu_e\rightarrow\nu_e}\simeq\cos^4\theta_{13}\sin^2\theta_{12}\cos^2\Big(\frac{\Delta{m}^2_2}{4E}L\Big)+\frac{1}{2}\sin^4\theta_{13}(1+\sin^22\theta_3)
\label{m_Ps22}
\end{eqnarray}
for $E>$ few MeV (such as $^8$B and $hep$ neutrinos). As expected, for the earth-sun distance $L_{\rm es}$ being much smaller than the oscillation length, {\it i.e.} $L_{es}\ll4\pi E/\Delta{m}^2_2$, we can obtain a similar form to the $3\nu$SF, $P^{m3\nu{\rm SF}}_{\nu_e\rightarrow\nu_e}\simeq\cos^4\theta_{13}\sin^2\theta_{12}+\sin^4\theta_{13}$. Consequently, for $^8$B neutrinos with energies above a few MeV, the SK+SNO\,\cite{Fogli:2003vj} and Borexino\,\cite{Agostini:2017ixy} data can explain the MSW-LMA solution to solar neutrino oscillations. For $L_{es}\gg4\pi E/\Delta{m}^2_2$ in Eq.\,(\ref{m_Ps22}), however, the oscillating phase averages out and the LMA solution cannot be explained.

On the other hand, as indicated in Refs.\,\cite{Abe:2016nxk, Aharmim:2011vm}, the recent observation of low-energy $^8$B solar neutrino flux (as low as $\sim3.5$ MeV) by SK\,\cite{Abe:2016nxk} has marked the raise of the solar neutrino tension that is a discrepancy appearing in the $3\nu$SF, see Eq.\,(\ref{sol_kam}). 
 This discrepancy could be due to the oscillating phase of Eq.\,(\ref{m_Ps22}) that only affects solar neutrinos. Interestingly enough, such discrepancy appearing in the $3\nu$SF can be removed by the new oscillation effect without significantly modifying the MSW-LMA solution to solar neutrino oscillations.
For oscillation lengths $L^{\rm osc}_2=4\pi E/\Delta{m}^2_2$ being optimized to $L_{\rm es}$, {\it that is},
\begin{eqnarray}
L^{\rm osc}_2=2.48\times10^{8}\,{\rm km}\,\Big(\frac{E}{1{\rm MeV}}\Big)\Big(\frac{10^{-11}{\rm eV}^2}{\Delta m^2_{2}}\Big)\gtrsim L_{es}\,,
\label{m2con}
\end{eqnarray}
a bound of $\Delta m^2_2\lesssim1.7\times10^{-11}\,{\rm eV}^2$ is roughly derived for energies above $1$ MeV in order not to significantly modify the current LMA solution, as illustrated in Fig.\,\ref{Fig4}.
In the $3\nu$SF, the mass-squared difference from the $^8$B solar neutrinos at SK\,\cite{Abe:2016nxk} is somewhat lower than that from the reactor neutrino at KamLAND\,\cite{Eguchi:2002dm, Araki:2004mb} at less than $2\sigma$, as shown in Eq.\,(\ref{sol_kam}). On the other hand, in our new oscillation framework, the data from the $^8$B solar neutrino experiments can be well fitted with that of the KamLAND $\bar{\nu}_e$, as illustrated in Fig.\,\ref{Fig4}: we find, numerically, that a solution to the solar neutrino tension can happen roughly at  
\begin{eqnarray}
0.9\times10^{-11}{\rm eV}^2<\Delta{m}^2_2\lesssim1.7\times10^{-11}\,{\rm eV}^2\,,
\label{m2con}
\end{eqnarray}
with the global fit of mixing parameters of $3\nu$SF in Table-\ref{exp_nu}.
Note here that the above estimation is derived from the $3\sigma$ data (instead of the $2\sigma$ data) of Table-\ref{exp_nu}.

\item
If the matter parameter $A_{e\mu}$ at $\nu_e$ production is well below the resonant value $A^r_{12}$ in Eq.\,(\ref{reso_r}), {\it i.e.} $A^r_{12}\gg A_{e\mu}(x_0, E)$, the corresponding matter effects are negligible, leading to $\theta^m_{12}\rightarrow\theta_{12}$. 
For $L_{es}\gg L^{\rm osc}_1=4\pi E/\Delta m^2_1$ with $0.1\,{\rm MeV}\lesssim E\lesssim{\rm few}$ MeV, the oscillating phase averages out. Then, it leads to $P^m_{\nu_e\rightarrow\nu_e}\simeq\cos^4\theta_{12}/2+\sin^4\theta_{12}\cos^2\big(\frac{\Delta{m}^2_2}{4E}L\big)$ which cannot explain the Borexino $pp$, $^7$Be, and $pep$ data\,\cite{Agostini:2017ixy}.
In fact, since the oscillation length $L^{\rm osc}_1$ at the low energy range in Eq.\,(\ref{m_Ps}) can be sensitive to the earth-sun distance $L_{es}$, it should be at least one order of magnitude larger than $L_{es}$ in order for the $pp$, $^7$Be, and $pep$ data shown in the left plot of Fig.\,\ref{Fig4} to fit well:
\begin{eqnarray}
L^{\rm osc}_1=2.48\times10^{9}\,{\rm km}\,\Big(\frac{E}{0.1{\rm MeV}}\Big)\Big(\frac{10^{-13}{\rm eV}^2}{\Delta m^2_{1}}\Big)\gg L_{es}\,,
\label{m1con}
\end{eqnarray}
whose condition leads to a bound\,\footnote{This bound satisfies the Borexino $pp$ data\,\cite{Agostini:2017ixy}, as clearly shown in the right plot of Fig.\,\ref{Fig4}, and which is stronger by one order of magnitude than a bound $\Delta m^2_1<1.8\times10^{-12}\,{\rm eV}^2$ at $3\sigma$ in Ref.\,\cite{deGouvea:2009fp}.} of $\Delta m^2_1$
\begin{eqnarray}
|\Delta m^2_{1}|\lesssim10^{-13}\,{\rm eV}^2\,.
\label{m1con2}
\end{eqnarray}
Then, the $\nu_e$ survival probability at the exposed surface of the Earth is given by
\begin{eqnarray}
P^m_{\nu_e\rightarrow\nu_e}&\simeq&\cos^4\theta_{13}\Big\{1-\frac{1}{2}\sin^22\theta_{12}-\sin^4\theta_{12}\sin^2\Big(\frac{\Delta{m}^2_2}{4E}L\Big)\Big\}\nonumber\\
&+&\frac{1}{2}\sin^4\theta_{13}(1+\sin^22\theta_3)\,,
\label{m_Ps12}
\end{eqnarray}
for $E\lesssim{\rm few}$ MeV (such as $pp$, $^7$Be, $hep$, $pep$, and $^8$B neutrinos\,\cite{Serenelli:2011py, PDG}). In this case,
the oscillating phase of Eq.\,(\ref{m_Ps12}) can make new appreciable effects when $L^{\rm osc}_2\lesssim L_{es}$, see the cyan curves of Fig.\,\ref{Fig4}, for energies less than 1 MeV.

\item
If the matter parameter $A_{e\mu}$ at $\nu_e$ production is only slightly below the resonant value, $A^r_{12}\gtrsim A_{e\mu}(x_0, E)$ in Eq.\,(\ref{mat_12}), the neutrino does not cross a region with resonant density, but matter effects are sizable enough to modify the mixing. Whereas, in the case that $A_{e\mu}(x_0, E)> A^r_{12}$, the neutrino can cross the resonance.
For both cases with energies $E\gtrsim1$ MeV (such as $^8$B and $hep$ neutrinos\,\cite{Serenelli:2011py, PDG}) and $L_{es}\ll4\pi E/\Delta m^2_{1}$, the new oscillatory term containing $\Delta m^2_2$ can play a crucial role in modifying the solar $\nu_e$ transition rate while satisfying the solar neutrino data. The solar $\nu_e$ survival probability then can be described by 
\begin{eqnarray}
P^m_{\nu_e\rightarrow\nu_e}&\simeq&\cos^2\theta_{13}\cos^2\theta^m_{13}\Big\{\cos^2\theta^m_{12}\cos^2\theta_{12}+ \sin^2\theta^m_{12}\sin^2\theta_{12}\cos^2\Big(\frac{\Delta{m}^2_2}{4E}L\Big)\Big\}\nonumber\\
&+&\frac{1}{2}\sin^2\theta^m_{13}\sin^2\theta_{13}(1+\sin^22\theta_3)\,.
\label{m_Ps0}
\end{eqnarray}
As shown in Fig.\,\ref{Fig4}, the oscillating phase in Eq.\,(\ref{m_Ps0}) dramatically modifies the gray shaded band predicted by the conventional MSW effect in the $3\nu$SF at energies $\gtrsim1$MeV. 
\end{itemize}
The resulting energy and oscillating parameters $\Delta{m}^2_{1(2)}$ dependence of the survival probability of solar neutrinos are shown in
Fig.\,\ref{Fig4} (together with a compilation of data from solar experiments\,\cite{napp}) where $\theta_{1(2)}\approx0$ and $\theta_3=-1.28$ (whose sizable value could be a solution to SBL anomalies\,\cite{Ahn:2019jbm}) are taken, as well as the prediction of the MSW-LMA solution of the $3\nu$SF using the $1\sigma$ parameter values (gray band) and the best-fits (black dotted line) given in Table-\ref{exp_nu}; the solar best-fits of Eq.\,(\ref{sol_kam}) (gray line). In the left plot of Fig.\,\ref{Fig4} for $\Delta m^2_1=10^{-13}\,{\rm eV}^2$ and $\Delta m^2_2=1.2\times10^{-11}\,{\rm eV}^2$, the cyan band corresponds to the $1\sigma$ range of the mixing parameters in Table-\ref{exp_nu}, especially, $\Delta m^2_{\rm Sol}=7.39^{+0.21}_{-0.20}\times10^{-5}\,{\rm eV}^2$ and $\sin^2\theta_{12}=0.310^{+0.013}_{-0.012}$ with $\sin^2\theta_{13}=0.02240^{+0.00065}_{-0.00066}$ and the blue solid line for the global best-fits, while the red solid line corresponds to the KamLAND best-fits of Eq.\,(\ref{sol_kam}) and the black dashed line for the solar best-fits of Eq.\,(\ref{sol_kam}) with $\sin^2\theta_{13}=0.0219\pm0.0014$. In the right plot of Fig.\,\ref{Fig4},  
the cyan band ($1\sigma$ parameter values given in Table-\ref{exp_nu}), blue solid line (best-fits in Table-\ref{exp_nu}), red solid line (KamLAND best-fits of Eq.\,(\ref{sol_kam})), and black dashed line (solar best-fits of Eq.\,(\ref{sol_kam})) correspond to $\Delta m^2_1=10^{-13}\,{\rm eV}^2$ and $\Delta m^2_2=1.2\times10^{-12}\,{\rm eV}^2$, while the orange band ($1\sigma$ parameter values given in Table-\ref{exp_nu}), blue solid line (best-fits in Table-\ref{exp_nu}), red solid line (KamLAND best-fits of Eq.\,(\ref{sol_kam})), and black dashed line (solar best-fit of Eq.\,(\ref{sol_kam})) for $\Delta m^2_1=10^{-12}\,{\rm eV}^2$ and $\Delta m^2_2=1.2\times10^{-11}\,{\rm eV}^2$.

In order to see a chi-square fit in probability space, we use the following expression for $\chi^2$:
\begin{eqnarray}
\chi^2=\sum_i\frac{(P^{\rm observed}_i-P^{\rm theory}_i)^2}{\sigma^2_i}\,.
\end{eqnarray}
Here $i$ runs over the observed measurements ($pp, ^7{\rm Be}, pep, ^8{\rm B}$ neutrinos in probability space), $P^{\rm observed}_i$ and $\sigma_i$ are the central values and uncertainties corresponding to a set of data shown in Fig.\,\ref{Fig4}, and the theoretical prediction for $P^{\rm observed}_i$ is $P^{\rm theory}_i$ (for example, Eq.\,(\ref{m_Ps}) for ``MSW+Oscillation" effect).
In the case that $P^{\rm theory}_i$ is given by Eq.\,(\ref{m_Ps}) for MSW+Oscillation effect, there are five free parameters $E, \theta_{12}, \Delta m^2_{21}, \Delta m^2_1$, and $\Delta m^2_2$, while there are three free parameters  in the case that $P^{\rm theory}_i$ corresponds to the case of 3$\nu$SF (for $\Delta m^2_{1(2)}\rightarrow0$).
\begin{figure}[h]
\includegraphics[width=10.0cm]{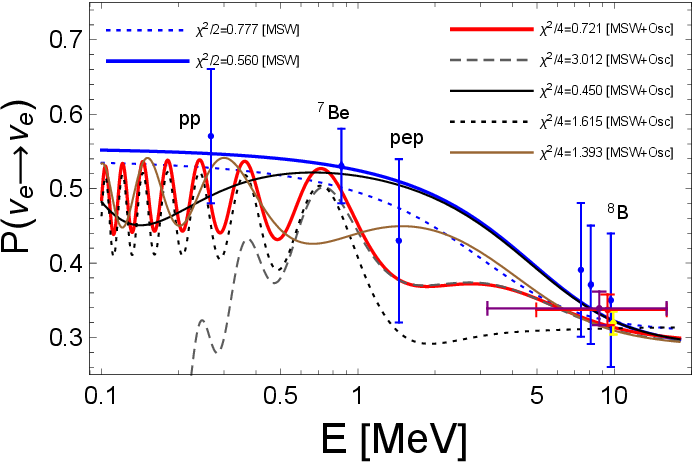}
\caption{\label{Fig5}Plot of $\chi^2$-fitting the model function of Eq.\,(\ref{m_Ps}) to a set of data. }
\end{figure}

Fitting the model function of Eq.\,(\ref{m_Ps}) to a set of data shown in Fig.\,\ref{Fig5}  via a chi square minimization,
we obtain the best-fit LMA for various parameters $\Delta m^2_{21}$ and $\Delta m^2_{1(2)}$ as shown in Table-\ref{chi_s}. In Fig.\,\ref{Fig5}, blue dotted (solar best-fit $\Delta m^2_\odot=4.85\times10^{-5}\,{\rm eV}^2$ of Eq.\,(\ref{sol_kam})) and blue solid (KamLAND best-fit $\Delta m^2_{\rm KL}=7.49\times10^{-5}\,{\rm eV}^2$ of Eq.\,(\ref{sol_kam})) lines are due to the conventional MSW effect of 3$\nu$SF, while red solid ($\Delta m^2_1=10^{-13}\,{\rm eV}^2$, $\Delta m^2_2=1.2\times10^{-11}\,{\rm eV}^2$), gray dashed ($\Delta m^2_1=10^{-12}\,{\rm eV}^2$, $\Delta m^2_2=1.2\times10^{-11}\,{\rm eV}^2$), and black solid ($\Delta m^2_1=10^{-13}\,{\rm eV}^2$, $\Delta m^2_2=1.2\times10^{-12}\,{\rm eV}^2$) lines are due to the composite matter effect ``MSW+Oscillation" with a given global best-fit $\Delta m^2_{\rm GF}$ in Table-\ref{exp_nu}. And black dotted (solar best-fit $\Delta m^2_\odot=4.85\times10^{-5}\,{\rm eV}^2$ of Eq.\,(\ref{sol_kam}) and $\Delta m^2_1=10^{-13}\,{\rm eV}^2$, $\Delta m^2_2=1.2\times10^{-11}\,{\rm eV}^2$) and brown solid line ($\Delta m^2_{21}=6.11\times10^{-5}\,{\rm eV}^2$ and $\Delta m^2_1=10^{-13}\,{\rm eV}^2$, $\Delta m^2_2=5\times10^{-12}\,{\rm eV}^2$) are due to the composite matter effect.
\begin{table}[h]
\caption{\label{chi_s} Chi-square fit of  $\theta_{12}$ for various $\Delta m^2_{21}$ and $\Delta m^2_{1(2)}$.}
\begin{ruledtabular}
\begin{tabular}{cccccccccc}
Model function &$\theta_{12}[^{\circ}]$&$\chi^2/{\rm d.o.f}$&goodness-of-fit&\vline\vline&$\Delta m^{2}_{21}[10^{-5}{\rm eV}^{2}]$&$\Delta m^2_1[{\rm eV}^{2}]$&$\Delta m^2_2[{\rm eV}^{2}]$\\
\hline
$P^{m(3\nu{\rm SF})}_{\nu_e\rightarrow\nu_e}$&$34.62$&$1.55/2$&54.0\%&\vline\vline&$4.85$&$0$&$0$\\
$P^{m(3\nu{\rm SF})}_{\nu_e\rightarrow\nu_e}$&$32.26$&$1.12/2$&57.1\%&\vline\vline&$7.49$&$0$&$0$\\
$P^m_{\nu_e\rightarrow\nu_e}$&$33.76$&$2.88/4$&57.8\%&\vline\vline&$7.39$&$10^{-13}$&$1.2\times10^{-11}$\\
$P^m_{\nu_e\rightarrow\nu_e}$&$33.48$&$12.05/4$&1.7\%&\vline\vline&$7.39$&$10^{-12}$&$1.2\times10^{-11}$\\
$P^m_{\nu_e\rightarrow\nu_e}$&$33.12$&$1.80/4$&77.2\%&\vline\vline&$7.39$&$10^{-13}$&$1.2\times10^{-12}$\\
$P^m_{\nu_e\rightarrow\nu_e}$&$35.07$&$6.46/4$&16.7\%&\vline\vline&$4.85$&$10^{-13}$&$1.2\times10^{-11}$\\
$P^m_{\nu_e\rightarrow\nu_e}$&$33.35$&$5.57/4$&23.3\%&\vline\vline&$6.11$&$10^{-13}$&$5\times10^{-12}$\\
\end{tabular}
\end{ruledtabular}
\end{table}

According to Table\,\ref{chi_s} and Fig\,\ref{Fig5}, the gray dashed line (with $\chi^2/{\rm d.o.f}=12.05/4$) is excluded at the $98.3\%$ C.L. Thus, a range of $\Delta m^2_1$ favors $\Delta m^2_1\lesssim10^{-13}\,{\rm eV}^2$ which is consistent with that of Eq.\,(\ref{m1con2}). As expected, the black solid line (with $\chi^2/{\rm d.o.f}=1.80/4$) will approach to the blue solid line  of 3$\nu$SF but give a much better fit due to a relatively large degrees of freedom when $\Delta m^2_2\lesssim1.2\times10^{-12}\,{\rm eV}^2$ leading to the vanishing oscillation effect. Thus, we can conservatively take a range of $\Delta m^2_2$ to be $\Delta m^2_2\simeq{\cal O}(10^{-11}\,{\rm eV}^2)$, as expected in Eq.\,(\ref{m2con}), in a similar ball park.
Comparing $\chi^2$ goodness-of-fit between the blue dotted line (with $\chi^2/{\rm d.o.f}=1.55/2$) and the blue solid line (with $\chi^2/{\rm d.o.f}=1.12/2$) for the MSW effect shows that the latter gives a better fit but comparable, however, the former may give a better but comparable fit once the data of Ref.\,\cite{Abe:2016nxk} are used. 
Interestingly enough, the $\chi^2$ goodness-of-fit of the red solid line (with $\chi^2/{\rm d.o.f}=2.88/4$) for the composite matter effect is well consistent with that of the blue solid line (with $\chi^2/{\rm d.o.f}=1.12/2$) for the only MSW effect.

All the data shown in Fig.\,\ref{Fig4} and \ref{Fig5}  that are consistent with the rough estimations of Eqs.\,(\ref{m2con}) and (\ref{m1con2}) favor
\begin{eqnarray}
\Delta m^2_1\lesssim10^{-13}\,{\rm eV}^2\,,\qquad\Delta m^2_2\simeq{\cal O}(10^{-11})\,{\rm eV}^2\,,
\end{eqnarray}
through the composite matter effects. This indicates that our new oscillation scenario can be a good candidate for explanation of a MSW-LMA solution to the solar neutrino tension.
Future precise measurements of $^8$B and $pep$ solar neutrinos may confirm and/or improve the value of $\Delta{m}^2_2$ as a solution to the solar neutrino tension, including future measurements of $hep$ solar neutrino which has not been detected yet. Moreover, future precise measurements of the carbon-nitrogen-oxygen (CNO) cycle neutrinos together with the recent measurements by Borexino Collaboration\,\cite{BoXi}, one of two sets of nuclear fusion reactions, will give a full understanding of solar neutrinos at less than few MeV, as well as the nuclear fission processes inside the Sun.

\section{Conclusion}
This is the first theoretical study of a would-be solution to the so-called solar neutrino tension why solar neutrinos at SNO, SK, and Borexino experiments appear to mix differently from reactor antineutrinos at KamLAND.
Three gauge-singlet neutrinos added to the standard model Lagrangian make the neutrinos massive, as required by experimental observations. 
A unitary condition is imposed to the $6\times6$ mixing matrix which connects the interaction eigenstates with the mass eigenstates.
Then the extended theory with three light sterile neutrinos forms pseudo-Dirac pairs that augment three additional oscillation parameter sets ($\Delta{m}^2_i$, $\theta_i$) besides the six oscillation parameters of the $3\nu$SF ($\Delta{m}^2_{\rm Sol}, \Delta{m}^2_{\rm Atm}$, $\theta_{12}, \theta_{23}, \theta_{13}$, $\delta_{CP}$): two $\Delta m^2_{\rm ABL}\lesssim{\cal O}(10^{-11})\,{\rm eV}^2$ optimized at ABL ($\gtrsim L_{es}=149.6\times10^6$ km, earth-sun distance) oscillation experiments and one $\Delta m^2_{\rm SBL}\sim{\cal O}(1)\,{\rm eV}^2$ optimized at reactor SBL oscillation experiments (with their corresponding mixing angles $|\theta_{1(2)}|\approx0\ll|\theta_3|\sim{\cal O}(1)$). 
If the light sterile neutrinos exist and have particular masses, each of them should produce a unique feature that is detectable by its optimized experiment. 

In the extended theory, we have derived a general transition probability between the massive neutrinos (that a flavor eigenstate $\nu_\alpha$ becomes flavor eigenstate $\nu_\beta$ with $\alpha,\beta=e,\mu,\tau$) that can have a potential for explaining the anomalous phenomena (the solar neutrino tension plus SBL anomalies) in terms of neutrino oscillations.
Assuming no sterile neutrinos are initially generated when electron neutrinos are produced in the Sun by nuclear reactions. 
Then, we have re-examined the MSW matter effects in our theoretical framework and suggested a solution to the solar neutrino tension with a so-called composite matter effect that causes a neutrino flavor change with new oscillatory terms containing $\Delta m^2_{\rm ABL}\lesssim{\cal O}(10^{-11})\,{\rm eV}^2$, so that $|\Delta m^2_1|\ll|\Delta m^2_2|\lesssim{\cal O}(10^{-11})\,{\rm eV}^2\ll|\Delta m^2_3|\sim{\cal O}(1)\,{\rm eV}^2$. We stress that, contrary to the conventional matter effect that causes a change in the flavor content of a neutrino but without sinusoidal oscillation, the composite matter effect causes a neutrino flavor change via the effects of sinusoidal oscillation induced by the oscillatory terms containing $\Delta m^2_{\rm ABL}$, as well as the MSW matter effect.

With the composite matter effect of our theoretical framework, we have shown that the values of $\Delta m^2$ measured in reactor KamLAND, $\Delta m^2_{\rm KL}=7.49^{+0.19}_{-0.18}\times10^{-5}\,{\rm eV}^2$\,\cite{Gando:2013nba}, can be compatible with those measured in solar neutrino experiments (SNO, SK, and Borexino) at energies ($>3$ MeV) for $\Delta m^2_1\lesssim10^{-13}\,{\rm eV}^2$ and $\Delta m^2_2\simeq{\cal O}(10^{-11})\,{\rm eV}^2$, as summarized in Fig.\,\ref{Fig4} and Fig.\,\ref{Fig5}. This indicates that our new oscillation scenario can be a good candidate for explanation of a MSW-LMA solution to the solar neutrino tension. However, as shown in Table\,\ref{chi_s} and Fig\,\ref{Fig5}, it is clear that the new scenario is not much more preferred than the standard case. In other words, the current data (solar data alone) is not precise enough to test the proposed scenario.
Future precise measurements of $^8$B and $pep$ solar neutrinos may confirm and/or improve the value of $\Delta{m}^2_2$ as a solution to the solar neutrino tension, including future measurements of $hep$ solar neutrino which has not been detected yet.

\acknowledgments{We would like to give thanks to Hyun Min Lee for useful discussions. This work was supported by Basic Science Research Program through the National Research Foundation of Korea (NRF) funded by the Ministry of Education, Science and Technology (NRF-2020R1A2C1010617) and (NRF-2019R1A2C2003738).
}

\end{document}